\documentclass[reprint,superscriptaddress,amsmath,amssym,aps,prb]{revtex4-2}

\usepackage{bm}
\usepackage{float}
\usepackage{graphicx}
\usepackage{mathtools}
\usepackage{braket}
\usepackage[usenames,dvipsnames]{color}
\usepackage[normalem]{ulem}
\usepackage[svgnames]{xcolor}
\usepackage{bm}
\usepackage{multirow}
\usepackage{titlesec}
\usepackage[utf8]{inputenc}
\usepackage{array}
\usepackage[colorlinks,linkcolor=blue,citecolor=blue,urlcolor=blue]{hyperref}

\begin{document}
\title{Scalable Sondheimer oscillations driven by commensurability between two quantizations}

\author{Xiaodong Guo}
\affiliation{Wuhan National High Magnetic Field Center and School of Physics, Huazhong University of Science and Technology,  Wuhan  430074, China}
\affiliation{Laboratoire de Physique et d'Etude de Mat\'{e}riaux (CNRS)\\ ESPCI Paris, PSL Research University, 75005 Paris, France }

\author{Xiaokang Li}
\affiliation{Wuhan National High Magnetic Field Center and School of Physics, Huazhong University of Science and Technology,  Wuhan  430074, China}

\author{Lingxiao Zhao}
\affiliation{Quantum Science Center of Guangdong-Hong Kong-Macao Greater Bay Area, Shenzhen 523335, China}

\author{ Zengwei Zhu}
\email{zengwei.zhu@hust.edu.cn}
\affiliation{Wuhan National High Magnetic Field Center and School of Physics, Huazhong University of Science and Technology,  Wuhan  430074, China}

\author{Kamran Behnia} 
\email{kamran.behnia@espci.fr}
\affiliation{Laboratoire de Physique et d'Etude de Mat\'{e}riaux (CNRS)\\ ESPCI Paris, PSL Research University, 75005 Paris, France }
\begin{abstract}
The electrical conductivity of metallic crystals exhibits size effects when the electron mean free path exceeds the sample thickness. One such phenomenon, known as Sondheimer oscillations, was discovered decades ago. These oscillations, periodic in magnetic field, have been hitherto treated with no reference to Landau quantization.  Here, we present a study of longitudinal and transverse conductivity in cadmium single crystals with thicknesses ranging from 12.6 to 475 $\mu$m, and demonstrate that the amplitude of the first ten oscillations is determined by the quantum of conductance and a length scale that depends on the sample thickness, the magnetic length and the Fermi surface geometry. We argue that this scaling is unexpected in semiclassical scenarios and it arises from the degeneracy of the momentum derivative of the cross-sectional area $A$ along the orientation of the magnetic field $\frac{\partial A}{\partial k_z}$ in cadmium, which couples Landau quantization to the discretization of $k_z$ imposed by the finite sample thickness. We show that the oscillating component of the conductivity is uniquely governed by fundamental constants and the ratio of two degeneracies, which acts as an inverted filling factor. Our conjecture is supported by the absence of such scaling in thin copper crystals. 
\end{abstract}
\maketitle

\section{Introduction}
Confinement of electrons to small spatial dimensions is known to affect electrical transport properties \cite{Sondheimer1952,BRANDLI196961,Chambers1971,Triverdi1988}.  Sondheimer discovered one such size effect \cite{Sondheimer1950}, which occurs when the sample thickness becomes shorter than the bulk mean free path. These oscillations of conductivity, sometimes dubbed magneto-morphic, are periodic in the magnetic field. Available explanations of Sondheimer oscillations (SO) do not invoke Landau quantization, in contrast to Shubnikov-de Haas oscillations. The latter are periodic in the inverse of the magnetic field \cite{Shoenberg_1984} and are known to arise from commensurability in momentum space between the Landau tubes and the extremal areas of the Fermi surface. The prevailing understanding of Sondheimer oscillations invokes commensurability between the orbits of the charged carriers in real space and the sample thickness\cite{hurd2012hall}. The phenomenon is classified as a semiclassical size effect, where neither the discrete energy levels introduced by the magnetic field nor those resulting from spatial confinement play a role.

\begin{figure*}[hbt!]
\centering
\includegraphics[width=1\linewidth]{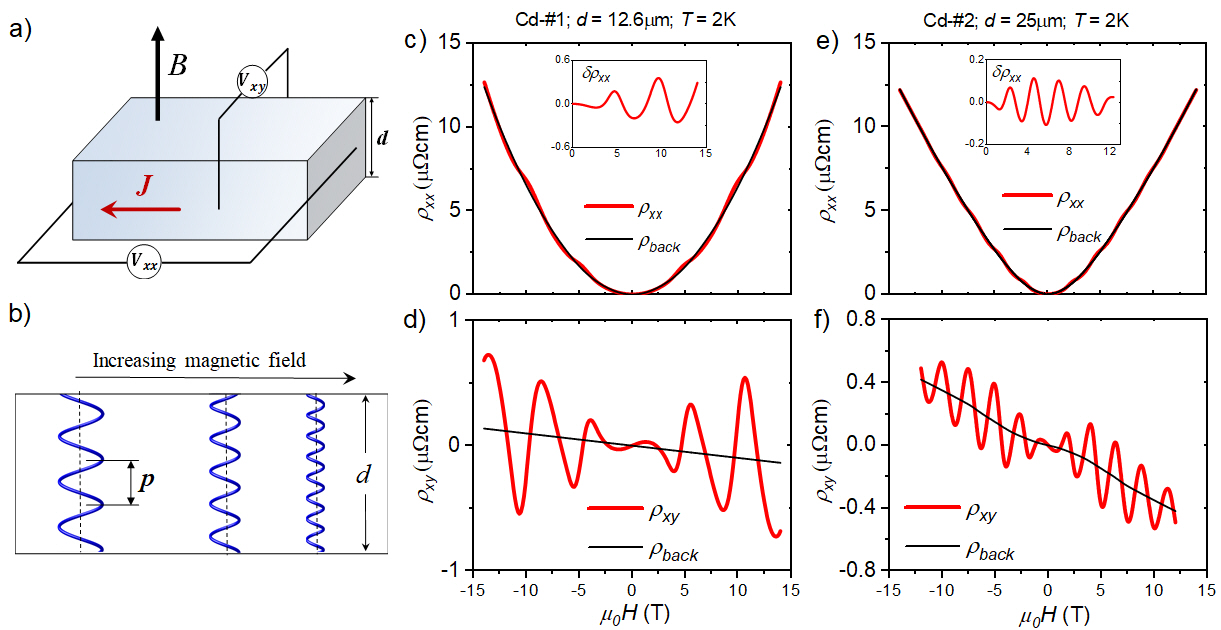} 
\caption{\textbf{Sondheimer oscillations and their extraction in two thin cadmium crystals.}
\textbf{(a)} Experimental setup. Electrons in a metallic solid conduct electricity in the presence of a magnetic field perpendicular to the current direction. Magnetoresistivity and Hall signals were measured in longitudinal and transverse configurations. 
\textbf{(b)} Front view of electron trajectories.
\textbf{(c)} Field dependence of longitudinal resistivity ($\rho_{xx}$) in the 12.6 $\mu$m sample. Red curve: experimental data; black line: non-oscillatory background. Inset: Oscillatory component after background subtraction.
\textbf{(d)} Hall resistivity ($\rho_{xy}$) of the 12.6 $\mu$m sample.
\textbf{(e)} Field dependence of the longitudinal resistivity ($\rho_{xx}$) in the 25 $\mu$m sample. Red curve: experimental data; black line: non-oscillatory background. Inset: Oscillatory component after background subtraction.
\textbf{(f)} Hall resistivity ($\rho_{xy}$) of the 25 $\mu$m sample.}
\label{fig.1}
\end{figure*}

Consider a magnetic field perpendicular to an electric current generating an electric field (Fig. \ref{fig.1}a). Owing to their Fermi velocity, mobile electrons experience a Lorentz force, giving rise to a helical trajectory (Fig. \ref{fig.1}b) whose axis is along the magnetic field. With increasing magnetic field, the radius of each turn of the helix decreases. Concomitantly, the distance between successive turns (i.e. the helix pitch, $p$) is reduced. At particular values of the magnetic field occurring with a fixed period, the thickness ($d$) of the sample becomes an integer multiple of helix pitches. In ballistic samples, where the mean free path limited by disorder is significantly longer than $d$, this condition leads to oscillations in conductivity, as observed in numerous experiments during the second half of the last century \cite{Grenier-cd,Al-2,Al-3,CD-2,Cd-3,Cd-4Zn-1,Trodahl_1971,Sakamoto1976,Munarin1968}. More recently, two studies have been devoted to Sondheimer oscillations in the Weyl semi-metal WP$_2$ \cite{van2021sondheimer} and in graphite \cite{Taen}.

The periodicity of the oscillations can be quantified thanks to a simple expression put forward by Harrison \cite{Harrison} for velocity:
\begin{equation}
v_z = \frac{\hbar}{2\pi m^*}\frac{\partial A}{\partial k_{z}}
 \label{1}
\end{equation}
Here, $v_z$ and $k_z$ represent the Fermi velocity and the wavevector along the magnetic field, respectively. $A$ is the cross-sectional area of the Fermi surface in the momentum space perpendicular to the magnetic field. The time for an electron to travel across the sample thickness is $\tau_1=d/v_z$. One revolution of the helix takes a time equal to $\tau_2=  \frac{2\pi m^\star}{eB}$. The ratio $\tau_1/\tau_2$ increases linearly with magnetic field and periodically reaches an integer. At these specific fields, the thickness equals an integer multiple of the helical pitch. The period, $\Delta B$, is therefore given by \cite{Chambers1971,Grenier-cd,SONDER-1,hurd2012hall,van2021sondheimer}:

\begin{equation}
\Delta B d = \frac{\hbar}{e} \frac{\partial A}{\partial k_{z}}.
 \label{2}
\end{equation}

\begin{table*}[hbt!] 
\begin{center} 
\renewcommand{\arraystretch}{1.3}
\begin{tabular}{|m{1.3cm}<{\centering}|m{3.5cm}<{\centering}|m{1.8cm}<{\centering}|m{1.8cm}<{\centering}|m{1.9cm}<{\centering}|m{2.8cm}<{\centering}|m{1.5cm}<{\centering}|m{2.1cm}<{\centering}|}   
\hline   \textbf{Sample} & \textbf{$d\times w \times l$ ($\mu$m$^3$)} & \textbf{$\Delta B_{MR}$ (T)} & \textbf{$\Delta B_{Hall}$ (T) } & \textbf{$\rho_0$ (n$\Omega\cdot$cm)}& \textbf{$\overline{s}=\sqrt{w \times d}$ ($\mu$m )} & \textbf{$\ell_0$ ($\mu$m)}  & \textbf{$\mu$ (m$^2$/V$\cdot$s)}  \\   
\hline   $\#$1 & (12.6 $\pm$ 0.8) $\times$ 24 $\times$ 47 & 5.04 & 5.14 & 7.37& 19.7 & 14.8 & 3.2 \\ 
\hline   $\#$2 & (25 $\pm$ 2)  $\times$ 23 $\times$ 49 & 2.33 & 2.4 & 6.23& 23.9 & 18.4 & 4.2 \\  
\hline   $\#$3 & (97.5 $\pm$ 3) $\times$ 160 $\times$ 298 & 0.58 & 0.58 & 2.9& 124 & 39 & 12 \\  
\hline   $\#$4 & (200 $\pm$ 8)  $\times$ 170 $\times$ 410 & 0.27 & 0.28& 1.42& 180 & 80.9& 20 \\ 
\hline   $\#$5 & (475 $\pm$ 10) $\times$ 362 $\times$ 420 & 0.12 & 0.12 & 1.45& 414 & 79.3 & 30\\
\hline 
\end{tabular}   
\end{center}
\caption{\textbf{Samples used in this study.} The field was oriented along the [0001] axis in all crystals. $\Delta B_{MR}$, $\Delta B_{Hall}$ represent the oscillation periods extracted from the magnetoresistance and the Hall resistivity. $\rho_0$ is zero-field resistivity measured at 2 K. The mean free path $\ell_0$ was extracted as described in the supplementary note 2 \cite{SM}. In the thickest samples, the zero-field electron mean free path $\ell_0$, limited by defect scattering, saturates at 75 $\mu$m; while in thinner samples, it is limited by boundary scattering, resulting in $\ell_0 \simeq \overline{s}$. }
\label{table:1} 
\end{table*}

In general, $\partial A /\partial k_{z}$ is not constant but varies with $k_z$. Therefore, a well-defined periodicity implies a singular value prevailing over the others \cite{gurevich1959oscillations,Chambers1971,hurd2012hall,abrikosov2017fundamentals}. In a spherical or an ellipsoidal Fermi surface, this singularity occurs at the end point of the Fermi surface along the orientation of the magnetic field. A distinct type of singularity occurs when $\partial^2A / \partial k_{z}^2 =0$ at an inflection point on the Fermi surface. The field dependence of the amplitude of oscillations is expected to depend on the type of singularity. In the case of an end point, the amplitude should fall off as $\propto B^{-4}$ \cite{Chambers1971}. This is what has been reported in cadmium and zinc \cite{Grenier-cd,CD-2,Cd-3,Cd-4Zn-1,Cd-5,cd-6,cD-7,cD-8,cD-9,cd-10}. In the case of an inflection point, one expects a field dependence $\propto B^{-2.5}$ \cite{gurevich1959oscillations}. This behavior has been observed in aluminum \cite{Al-2,Al-3}. Hurd \cite{hurd2012hall} has proposed that other possible exponents for the power law decay can arise for other types of  $\partial A /\partial k_{z}$  singularity.

\begin{figure*}[hbt!]
\centering
\includegraphics[width=1.0\linewidth]{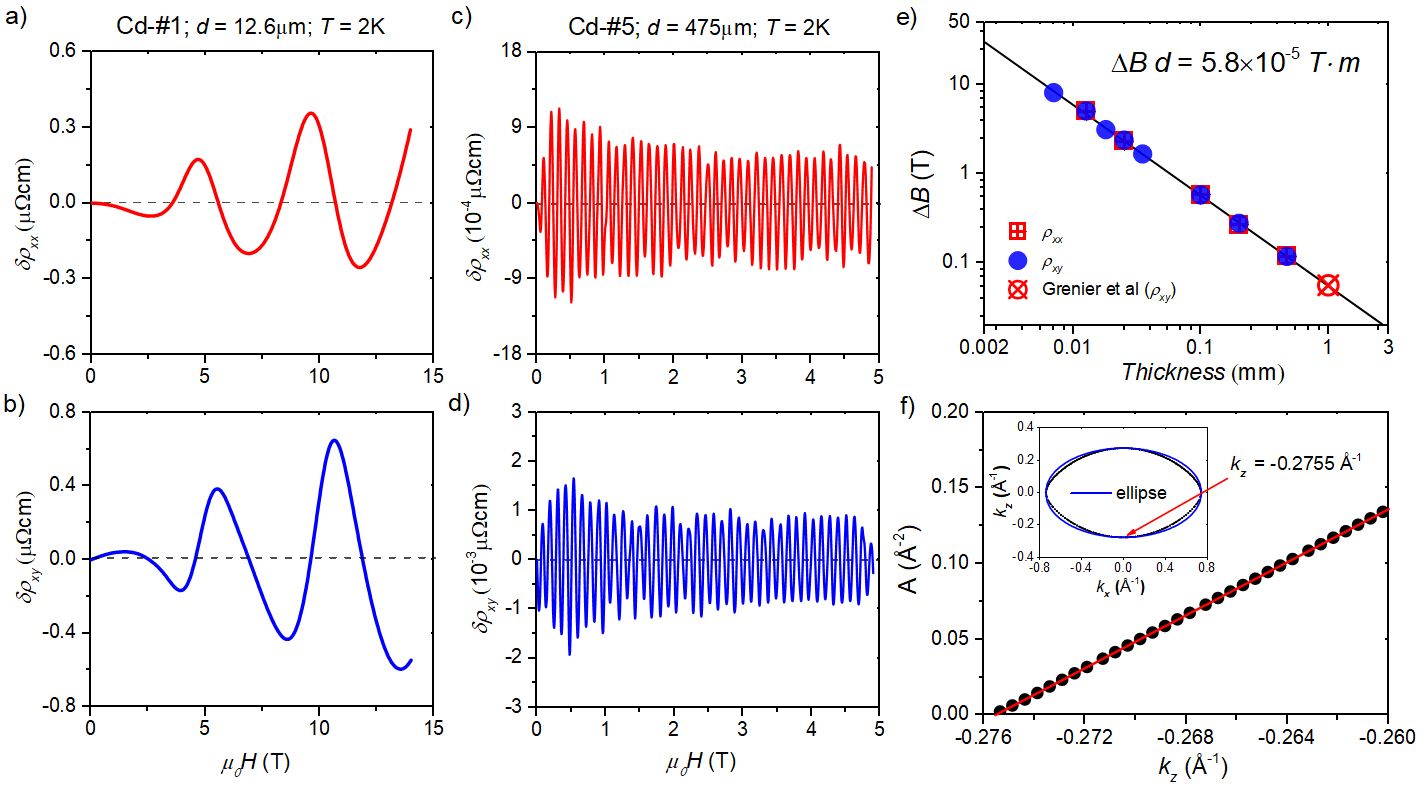} 
\caption{\textbf{Thickness dependence of oscillation period and Fermi surface topology.} \textbf{(a)},\textbf{(b)} The amplitude of Sondheimer oscillations (SO) in the longitudinal and transverse resistivity for the 12.6 $\mu$m sample, respectively. As the field increases, the amplitude of oscillation grows. \textbf{(c)},\textbf{(d)} The amplitude of SO in the longitudinal and transverse resistivity for the 475 $\mu$m sample, respectively. \textbf{(e)}  Oscillation period, $\Delta B$, as a function of thickness, $d$. The solid line represents a linear variation with a slope corresponding to $\Delta B\cdot d = 58$ T$\cdot \mu$m. The data point reported by Grenier \textit{et al.}\cite{Grenier-cd} for a 1.02 mm thick sample is also included. Inserting this $\Delta B d$ in Equation \ref{2} yields $\frac{\partial A}{\partial k_z}= 8.76$ \AA$^{-1}$. \textbf{(f)}  The computed  cross section of $A$ for the `lens-shaped' electron-like pocket as a function of $k_z$ shown near the southern pole \cite{Subedi2024}. This pocket, as shown in the inset, is not elliptical. The plot of $A$ \textit{vs.} $k_z$ is perfectly linear, and its slope yields  $\frac{\partial A}{\partial k_z}=\pm 8.73$ \AA$^{-1}$. This corresponds within a margin less than a percent to the experimentally derived  $\frac{\partial A}{\partial k_z}$.  The linear out-of-plane dispersion of this band, which hosts semi-Dirac fermions, is the origin of the constant $\frac{\partial A}{\partial k_z}$ over a $k_z$ interval of at least 0.014 \AA$^{-1}$. }
\label{fig.2}
\end{figure*}

\begin{figure*}[hbt!]
\centering
\includegraphics[width=1.0\linewidth]{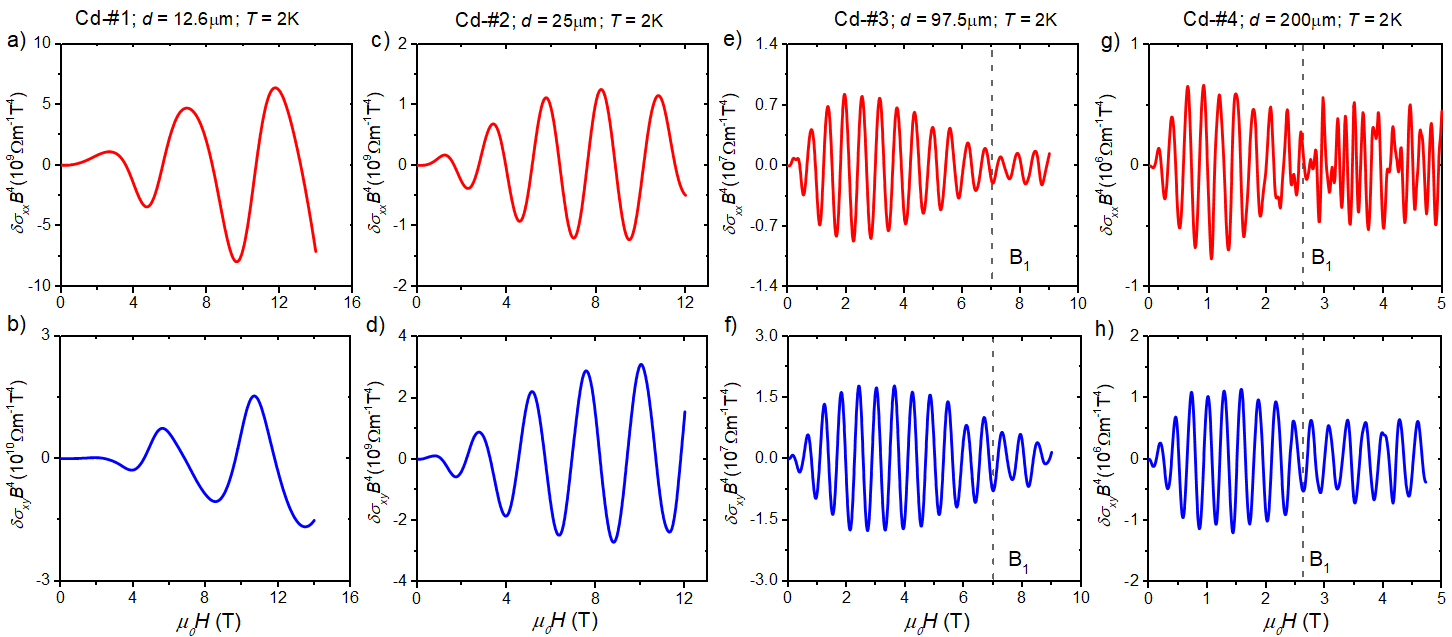} 
\caption{\textbf{Failure of the $B^{-4}$ field dependence.} \textbf{(a)},\textbf{(b)} Multiplying the longitudinal and transverse conductivity of the 12.6 $\mu$m sample by $B^{4}$ does not yield oscillations with a constant amplitude. \textbf{(c)},\textbf{(d)}  Same (a) and (b), but for the 25 $\mu$m sample.  \textbf{(e)},\textbf{(f)} Same (a) and (b), but for the 97.5 $\mu$m sample. A change in regime is observed above 7.1 T. \textbf{(g)},\textbf{(h)} For the 200 $\mu$m sample, the amplitude of $\delta \sigma_{xx}B^4$ oscillations becomes constant above 2.6 T.}
\label{fig.3}
\end{figure*}

\begin{figure*}[ht]
\centering
\includegraphics[width=1.0\linewidth]{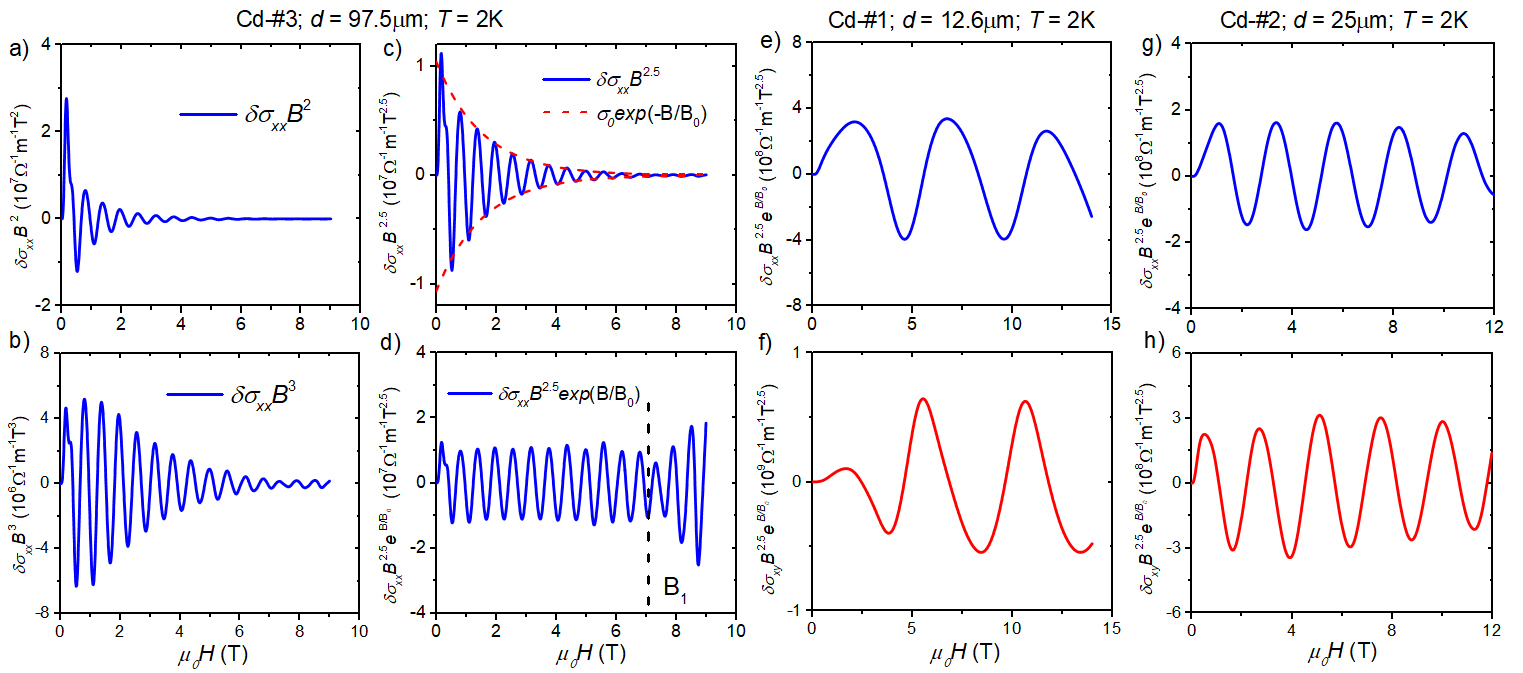} 
\caption{\textbf{Identification of the best-fit scaling: $\propto B^{-2.5}$$e^{-B/B_0}$.} \textbf{(a)} The longitudinal conductivity multiplied by $B^{2}$. The envelope of the oscillation amplitude decreases monotonically. \textbf{(b)} The longitudinal conductivity multiplied by $B^{3}$. The oscillation amplitude is non-constant and its dependence on field is non-monotonic. \textbf{(c)} The longitudinal conductivity multiplied by $B^{2.5}$, showing an exponential decay (indicated by a red dashed line). \textbf{(d)} Multiplying by $B^{2.5}$$e^{B/B_0}$ leads to a flat evolution of oscillation amplitude  with magnetic field. \textbf{(e)} The longitudinal conductivity of the 12.6 $\mu$m sample multiplied by  $B^{2.5} e^{B/B_0}$. Sinusoidal oscillations are observed with a period of 5.14 T. Here, the $B_0$ equals 15.2 T. \textbf{(f)} The same analysis for Hall conductivity yields a similar period $B_0$ approximately 16.3 T. \textbf{(g)}, \textbf{(h)} Application of the same method for the 25 $\mu$m sample reveals sinusoidal oscillations with a period of approximately 2.42 T. $B_0$ is 6.8 T and  7.2 T  for the longitudinal and transverse conductivity.}
\label{fig.4}
\end{figure*}

We present here a study of longitudinal and Hall resistivity of cadmium single crystals with varying thicknesses ($d$), ranging from 12.6 to 475 $\mu$m. Employing focused ion beam (FIB) technology \cite{Moll2018} to etch the samples, we ensured planar and parallel surfaces. Our extensive study on five different samples with a forty-fold variation in thickness has led to solid conclusions regarding the period and the amplitude of oscillations and their evolution with magnetic field and thickness. The period of oscillations, $\Delta B$, agrees with what was reported by Grenier \textit{et al.} \cite{Grenier-cd} for a thicker sample. In contrast, the amplitude of oscillations in thin samples, follows $\delta \sigma \propto B^{-2.5}e^{-B/3\Delta B}$. Such a field dependence was never proposed or observed before. Only when the samples become sufficiently thick, and the magnetic field sufficiently high, the field dependence becomes $\sigma \propto B^{-4}$ as previously reported. The crossover between the two regimes of field dependence occurs at  $\sim 10$ oscillations. 

Section II presents our extensive data sets leading to an empirical determination of the evolution of the oscillation amplitude as a function of magnetic field and thickness.  We show that for all crystals, the best fit is given by $\delta\sigma\propto B^{-2.5}e^{-B/B_0}$. Over a factor of  forty variation in thickness, the amplitude scales $\propto d^{-2}$. We also show that the temperature dependence of the oscillations is such that they vanish when the thermal energy exceeds the spacing between the Landau levels, implying a role played by the latter, which are not invoked in the semiclassical scenario. 

In section III, we contrast this distinct dependence on magnetic field and thickness with what is expected in the semiclassical models. We show that in our case, the data can be expressed as the oscillations of a dimensionless conductivity as a function of a dimensionless period.  We argue that the specific electronic structure of Cd \cite{Subedi2024} is crucial for this scalability. This conjecture is supported by a comparison between Cd and Cu. We then argue that the degeneracy of $\partial A /\partial k_{z}$ over a significant portion of the Fermi surface leads to an interplay between Landau levels and spatial confinement. The competition between two distinct discretizations of the energy spectrum, imposed by Landau quantization and $z$-axis confinement, generates oscillations whose period set by the commensurability between these two quantum degeneracies. The exponential term, unexpected in any of the available semiclassical pictures, can be tracked as the consequence of tunneling across two neighboring Landau levels. This scenario provides a basis for the expression empirically derived in section II.

\section{Results}

\subsection{Samples}

The crystal growth and the fabrication of the samples using FIB are discussed in detail in the supplement (Note 1) \cite{SM}. Table \ref{table:1} lists the samples used in this study.  Their thickness was varied by a factor of 40.  In the thickest samples, the zero-field resistivity at 2 K was found to be as low as $\sim1.5$ n$\Omega\cdot$cm, implying a mean free path as long as $\sim$80 $\mu$m.  This corresponds to the mean free path set by the distance between defects. In samples thinner than 80 $\mu$m, the residual resistivity increases because of the ballistic limit on the mean-free-path. In this regime, the mean-free-path extracted from the zero-field conductivity is comparable to the geometric average of thickness and width of the sample. 

\begin{figure}[bht!]
\includegraphics[width=1.0\linewidth]{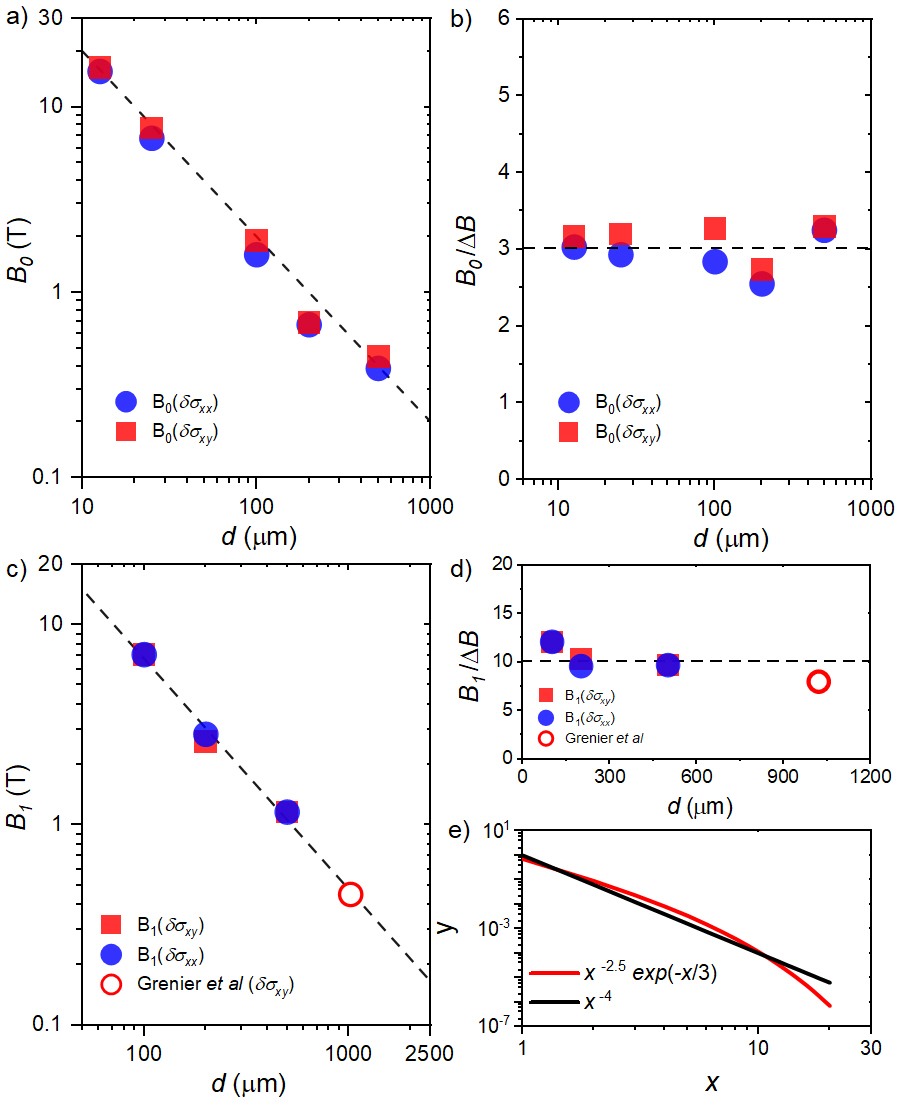} \caption{\textbf{Thickness dependence of $B_0$ and $B_1$.} \textbf{(a)} $B_0$ is inversely proportional to the thickness. \textbf{(b)} In all samples, $B_0$ is about 3 times $\Delta B$. \textbf{(c)} $B_1$ in different samples. It decreases with increasing $d$. The data for $d$ = 1.02 mm is taken from Ref. \cite{Grenier-cd}. \textbf{(d)} The ratio of $B_1/\Delta B$ is close to  10 in all samples. \textbf{(e)} A comparison of $g(x)= x^{-4}$ and $f(x)=$  exp$(-x/3)\cdot x^{-2.5}$. The two functions cross each other at $x^\star\simeq10.5$.}
\label{fig.5}
\end{figure}

\subsection{Extracting the oscillating component of the conductivity}
Fig. \ref{fig.1}\textbf{c} and \ref{fig.1}\textbf{d} show the data for our thinnest samples, with a thickness of $12.6~\mu$m.   As seen in Fig.\ref{fig.1}\textbf{c}, the longitudinal resistivity is dominated by a large background.  Cadmium is a compensated metal with a finite overlap between the conduction and the valence bands and an equal density of holes and electrons. Like other elemental semi-metals, such as bismuth \cite{Zhu_2018} and antimony \cite{Fauque2018}, it hosts a large orbital magnetoresistance, which, thanks to perfect compensation, does not saturate even in the high field limit ($\mu B \gg 1 $, where $\mu$ is the electronic mobility) \cite{pippard1989magnetoresistance}. On top of this background, one can detect oscillations. The inset shows the oscillating signal extracted by using a quadratic fit to the monotonic background. Two oscillations and the beginning of the third one are clearly visible. 

Fig. \ref{fig.1}\textbf{d} shows the Hall resistivity of the same sample. Here, one can see that field-periodic oscillations dominate the response, which is not surprising given the almost perfect compensation between the density of electrons and holes. The slope of the black solid line implies a finite non-oscillating Hall resistivity, presumably due to a slightly larger mobility of electrons compared to holes. 

Fig. \ref{fig.1}\textbf{e}  and Fig. \ref{fig.1}\textbf{f} show similar data for a twice thicker ($d=25~ \mu$m) sample. There are twice more oscillations in the same field window and the frequency of oscillations become twice faster. Meanwhile, the amplitude of oscillations is also reduced. The shape of oscillations is close to sinusoidal, but the evolution of the amplitude with magnetic field is complex.

Raw data for three thicker samples are presented in supplementary note 3 \cite{SM}. These data confirm the features seen in the thinnest samples.

\subsection{Period of oscillations}
Fig. \ref{fig.2} displays the evolution of the signal in two Cd single crystals with $d$ ranging from 12.6 to 475 $ \mu $m. Note the scales of both the vertical and horizontal axes vary across the panels. The absolute amplitude of oscillations and their period both enhance with thinning.  Since oscillations are squeezed in thicker samples, the field axis is limited to a lower bound to make them visible. More than one frequency may be present in the thickest ($d = 475~ \mu$m) sample. Here, we will focus on the main one. 

The evolution of the main period as a function of the thickness is shown in Fig. \ref{fig.2}\textbf{e}. One can see that our data for $10~\mu$m $< d < 500 ~\mu$m are in excellent agreement with what was found by Grenier \textit{et al.} \cite{Grenier-cd} for a sample with $d = 1020~ \mu$m. The solid black line represents a linear relation between thickness and frequency. The product of period and thickness is constant ($\Delta B \cdot d =$ 0.058 T$\cdot$mm) even when $d$ varies by two orders of magnitude. Inserting this in Equation \ref{2} leads to $\frac{\partial A}{\partial k_{z}}$ = 8.76 \AA$^{-1}$. This is to be compared with what is calculated by \textit{ab initio} theory $\frac{\partial A}{\partial k_{z}}$ = 8.73 \AA$^{-1}$ \cite{Subedi2024}.  The agreement is remarkable. 

The semi-Dirac nature of the band giving rise to the electron-like pocket plays an important role. This pocket is not an ellipsoid \cite{Grenier-cd, Subedi2024} (See the inset in Fig. \ref{fig.2}\textbf{f}).  In an ellipsoid, the extremal $\frac{\partial A}{\partial k_{z}}$ at each pole (or apex) is equal to $\frac{\partial A}{\partial k_{z}}=2\pi \frac{k_{Fr}^2}{k_{Fz}}$ = 12.6 \AA$^{-1}$, significantly different from the experimentally measured value (The computed pocket dimensions are $k_{Fr} = 0.742~$\AA$^{-1}$ and  $k_{Fz} = 0.2755~$\AA$^{-1}$).  Moreover, in contrast to an ellipsoid, a large portion of the Fermi surface near the poles shares an identical $\frac{\partial A}{\partial k_{z}}$. This can be seen in Fig. \ref{fig.2}\textbf{f}, which shows $A$ as a perfectly linear function of $k_z$ near a pole ($k_z^{max}=-0.2755$ \AA$^{-1}$).  This will play a crucial role in our discussion in section III.

\subsection{Amplitude of oscillations}
To compare the amplitude of oscillations with theoretical expectations, we inverted the measured resistivity tensor to quantify the conductivity tensor. Fig. \ref{fig.3} displays $\delta\sigma_{xx}B^{4}$ and $\delta\sigma_{xy}B^{4}$ as a function of magnetic field in four samples. By plotting the data in this way one can see that the postulated $\propto B^{-4}$ field dependence does not hold and $\sigma_{ij}B^{4}$ oscillation peaks do not keep a fixed amplitude. 

To empirically explore possible fits to the field dependence of the amplitude of conductivity oscillations, we multiplied the longitudinal conductivity by various powers of the magnetic field. Fig. \ref{fig.4}  shows the data for the 97.5 $\mu$m sample.  Plots of $\delta \sigma_{xx} B^\alpha$ (with $\alpha$ = 2, 2.5 and 3) are shown.  None of them are satisfactory. Upon closer examination, we found that  $\delta \sigma_{xx} B^\alpha$ becomes non-monotonic when $\alpha > 2.5$, but is monotonic when $\alpha <2.5$. Moreover, when $\alpha =2.5$,  $\delta \sigma_{xx} B^{2.5}$ displays an exponential decay (see Fig. \ref{fig.4}\textbf{c}, indicated by the red dashed line). This led us to try $\propto B^{-2.5}e^{-B/B_0}$ (see Fig. \ref{fig.4}\textbf{d}) with $B_0=1.8$ T.  Required for dimensional consistency, $B_0$ is a parameter indispensable for ensuring an almost perfect sinusoidal behavior. The deviation from this behavior seen in high fields (Fig. \ref{fig.4}\textbf{d}) is also meaningful.

This fitting function, $\propto B^{-2.5}e^{-B/B_0}$, was applied to the data obtained from other samples ($d < 100 ~ \mu$m). As illustrated in Fig. \ref{fig.4}, in samples with a thickness of 12.6 (Fig.  \ref{fig.4}\textbf{e, f}) and 25 (Fig. \ref{fig.4}\textbf{g, h}) $\mu$m, oscillations of conductivity (both longitudinal and transverse) are $\propto B^{-2.5} e^{-B/B_0}$ up to the highest magnetic field ($B <$ 14 T). The few oscillations are almost perfectly sinusoidal with no deviation occurring at high magnetic fields.  In the 200 and 475 $\mu$m samples, conductivities follow  $\propto B^{-2.5} e^{-B/B_0}$ below a threshold field, which we call $B_1$.  Above $B_1$, both conductivities follow  $\propto B^{-4}$ (see the supplementary note 4 \cite{SM}).

\begin{figure*}[ht!]
\includegraphics[width=0.9\linewidth]{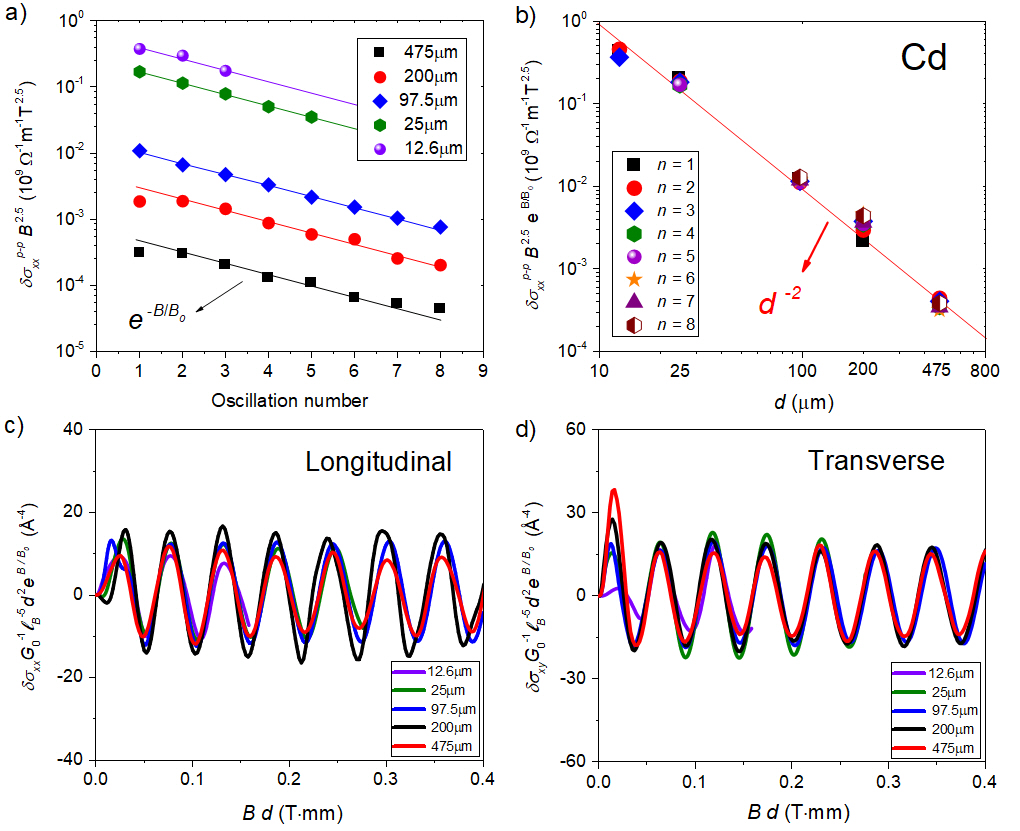} 
\caption{\textbf{Thickness dependence and scaling of the
amplitude.} \textbf{(a)} The peak-to-peak amplitude of the
oscillations, $\delta \sigma_{xx}^{pp} B^{2.5}$, evolves with oscillation number and exhibits an exponential decay. \textbf{(b)} The peak-to-peak amplitude of the oscillations, $\delta \sigma_{xx}^{pp} B^{2.5}e ^{B/B_0}$, scales with thickness $d$. The best fit is close to $d^{-2}$. \textbf{(c)} Oscillations of normalized conductivity as a function of the product of magnetic field and thickness ($B d$). Curves for all samples fall on top of each other. The vertical axis represents conductivity divided by the conductance quantum ($G_0=\frac{2e^2}{h}$) and multiplied by $\ell_B^{-5} d^2 e^{B/B_0}$. This normalized quantity has dimensions of [$L^{-4}$]. \textbf{(d)} The same scaling applied to the Hall conductivity data. All curves collapse, with normalized amplitude twice that of the longitudinal component.} 
\label{fig.6}
\end{figure*}

Fig. \ref{fig.5}\textbf{a} shows the thickness dependence of $B_0$. We find $B_0 \propto d^{-1}$. As mentioned above, the period of oscillations is also proportional to $d^{-1}$. As a result, the ratio of $B_0$ to $\Delta B$ is constant, as revealed in Fig. \ref{fig.5}\textbf{b}. Even more striking is the fact that in five samples studied,  the ratio is close to three: $\frac{B_0}{\Delta B}\sim 3$. Thus, $B_0$ is not an additional parameter, but is simply three times $\Delta B$. To summarize our results, the field dependence of the (peak-to-peak) amplitude of oscillations has two distinct regimes: 
\begin{equation}
\delta \sigma^{pp} \propto \left\{
             \begin{array}{lr}
             B^{-2.5}e^{-B/3 \Delta B}, & B < B_1.\\
             B^{-4}, & B > B_1. 
             \end{array}
\right.
\label{3}
\end{equation}
Let us now turn our attention to the significance of $B_1$.

\subsection{Boundary between the two regimes}
As the sample thickens, the threshold field $B_1$ decreases. This ensures consistency between our data and what was reported in the previous study of Sondheimer oscillations in cadmium, which was performed on a 1.02 mm thick sample \cite{Grenier-cd} and concluded that the amplitude of oscillations follows $B^{-4}$. Interestingly, the authors found a deviation from this behavior in their data below  $\sim 0.45$ T (See Fig.13 in ref. \cite{Grenier-cd}), in agreement with our findings.

As seen in Fig. \ref{fig.5}\textbf{c}, which shows the thickness dependence of $B_1$, the latter is proportional to the inverse of the thickness. The plot includes a data point for $d=1.02$ mm \cite{Grenier-cd} with  $B_1$  $\sim 0.45$ T. Fig. \ref{fig.5}\textbf{d} shows that for all samples, $\frac{B_1}{\Delta B}\simeq 10$. For more details on the boundary between the two regimes, see the supplementary note 4 \cite{SM}. 

Insight into this is offered by considering the two functions $g$ and $f$ of $x=B/\Delta B$. With $g(x)= x^{-4}$ and $f(x)=$exp$(-x/3)x^{-2.5}$, one can see that they cross each other at $x^\star\simeq10.5$. In other words, $f>g$ when $x<x^\star$ and  $f<g$ when $x>x^\star$. This means that the $B_1= 10 \Delta B$ is not an independent parameter, but a marker of domination between two competing field dependencies. $\Delta B$ is the fundamental field scale.

Thus, $\sigma_{ij}\propto B^{-2.5}e^{-B/3\Delta B}\cos(B/\Delta B)$  field dependence, which holds in thinner samples, coexists with the $\sigma_{ij}\propto B^{-4}\cos(B/\Delta B)$ field dependence. The latter dominates in thicker samples and high magnetic field, just because it decays slower at higher fields.   

Let us now turn our attention to the thickness dependence.

\subsection{Thickness dependence and the scaling of the amplitude}

Fig. \ref{fig.6}\textbf{a} is a plot of the product of $\delta \sigma^{pp}$ and $B^{2.5}$ as a function of oscillation number, confirming the relevance of the exponential term. Fig. \ref{fig.6}\textbf{b} shows how the peak-to-peak amplitude of the oscillations, $\delta \sigma^{pp}$, evolves with thickness. The product of $\delta \sigma^{pp}$ and $B^{2.5}e^{B/B_0}$ in each sample is plotted as a function of its thickness, $d$.  It is clear that the best fit is close to $d^{-2}$. As we will discuss in the next section, neither this thickness dependence nor this field dependence is expected in the available semiclassical theories. As seen in Fig. \ref{fig.6}\textbf{c}, \textbf{d}, this field and thickness dependence leads to a scaling of the amplitude of the oscillations (both longitudinal and transverse in all cadmium single crystals used in this study). All curves collapse on top of each other. 
 
Thus, not only the period but also the amplitude of the first ten oscillations is determined by the thickness and Fermi surface geometry.  This scaling implies that the amplitude can be written as:

\begin{equation}
\delta \sigma^{pp} = G_0 d^{-2} e^{-B/3\Delta B} \ell_B^{5} k_s^4
    \label{OSC}
\end{equation}
Here, $\ell_B=\sqrt{\frac{\hbar}{eB}}$ is the magnetic length. The parameter $k_s$ is a momentum space length scale, determined by the Fermi surface geometry.

\subsection{Temperature dependence of the oscillations} 
\begin{figure*}[ht]
\includegraphics[width=0.8\linewidth]{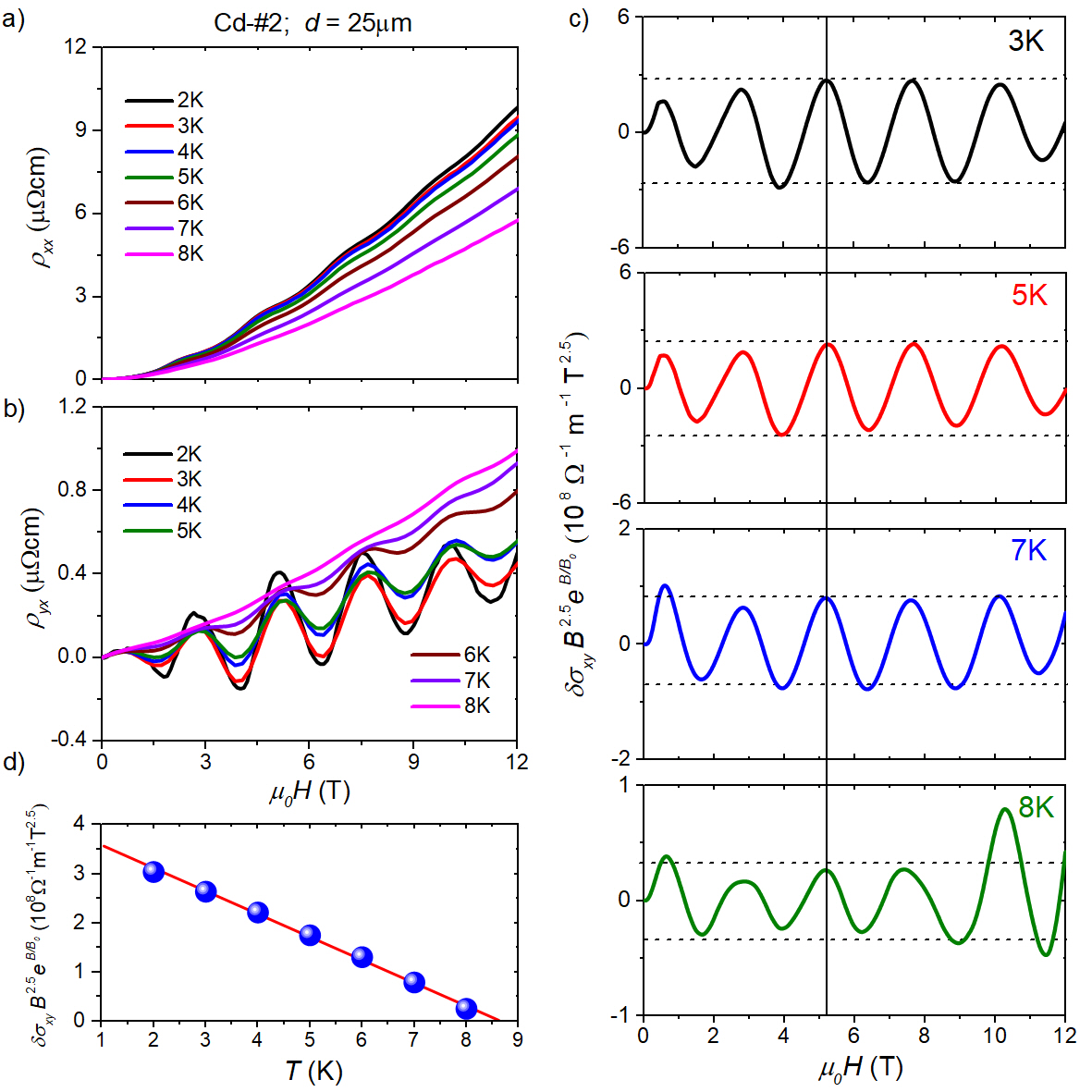} 
\caption{\textbf{Temperature dependence of oscillations in 25 $\mu$m sample.} 
\textbf{(a)} Longitudinal resistivity $\rho_{xx}$ as a function of magnetic field at different temperatures. \textbf{(b)} Transverse resistivity $\rho_{xy}$ as a function of magnetic field at different temperatures. \textbf{(c)} Transverse conductivity $\delta \sigma_{xy}$ multiplied by $B^{2.5} e^{B/B_0}$, showing the flattened evolution of the oscillation amplitude with magnetic field. \textbf{(d)} The amplitude of the third oscillation marked by the black line in \textbf{(c)} as a function of temperature.}
\label{fig.7}
\end{figure*}

Only the temperature dependence of resistivity in the 25 $\mu$m sample was carefully studied. As shown in Fig.\ref{fig.7}, magneto-morphic oscillations in both longitudinal and Hall resistivity weaken with increasing temperature. The evolution of $\delta\sigma_{xy}B^{2.5}e^{B/B_0}$ with temperature is shown in Fig. \ref{fig.7}\textbf{c}. These oscillations remain sinusoidal across all measured temperatures. As seen in Fig. \ref{fig.7}\textbf{d}, the amplitude of the oscillations decreases linearly with increasing temperature and vanishes when $T \simeq$ 8.5 K. 

Given that the effective mass of carriers is not far from the bare electron mass, in a magnetic field of $\sim 10 $ T, this is a temperature at which the thermal energy exceeds the spacing between Landau levels.

\section{Discussion}
Our extensive data analysis led to an empirical expression for oscillating conductivity \ref{OSC}. In this section, we begin by showing that this expression deviates from the available semiclassical theories. A comparison with copper indicates that specific features in the band structure of cadmium are essential. Then, we will show that, within a quantum picture, the oscillations arise because sweeping the magnetic field drives two distinct discretizations into commensurability at regular intervals of magnetic field, and the exponential term in equation \ref{OSC} can be traced to the mesoscopic degeneracy of $\frac{\partial A}{\partial k_{z}}$. Finally, we reformulate the empirical equation \ref{OSC} and render its connection to the Fermi surface geometry transparent.

\subsection{Comparison with semiclassical predictions} 

Abrikosov \cite{abrikosov2017fundamentals} considered the field dependence of Sondheimer oscillations in two distinct cases. In the first case, the Fermi surface has an inflection point (at which $\frac{\partial^2A}{\partial k_{z}^2}=0$). He found that in this case, the amplitude of oscillations decays with increasing magnetic field (following $B^{-5}$) and thickness (following $d^{-3/2}$). In other words:

\begin{equation}
\delta \sigma_{ip}^{pp} \propto \ell_B^{5}d^{-3/2}
    \label{inflection}
\end{equation}

This result was originally derived by Gurevich \cite{gurevich1959oscillations}. In addition, Abrikosov considered the case of an end point on a featureless Fermi surface and found a result similar to what was found by Chambers \cite{Chambers1971}. In the latter case, the amplitude of oscillations was found to be: 

\begin{equation}
\delta \sigma_{ep}^{pp} \propto \ell_B^{8}d^{-3}
    \label{endpoint}
\end{equation}

As shown in the supplementary Note 4 \cite{SM}, our high-field oscillations do not display a scaling as remarkable as that seen for low-field oscillations. However, their thickness and field dependence are roughly compatible with Eq. \ref{endpoint}. This is understandable given that there is indeed an end point in the Fermi surface of cadmium. As we will see below, the field dependence in copper is similar to what is expected by Eq. \ref{inflection}. 

Our low field empirical expression Eq. \ref{OSC}, however, differs from both Equations \ref{inflection} and \ref{endpoint}. The most striking difference is the presence of $e^{-B/3\Delta B}$; the thickness dependence also differs. There are other semiclassical scenarios, in which the exponent of the power law is different from -2.5 or -4 depending on Fermi surface details \cite{hurd2012hall}. However, none of these semiclassical scenarios account for the exponential term, which is reminiscent of the canonical quantum tunneling expression $e^{-s/\hbar}$, where $s$ represents an action \cite{WILKINSON1986341}.

\subsection{Contrasting cadmium with copper } 
To determine whether the scaling discovered in cadmium is general, we used FIB to fabricate thin copper crystals in a manner identical to that employed for cadmium. As discussed in the supplement (note 5) \cite{SM}, we detected Sondheimer oscillations in both longitudinal and Hall conductivities after subtracting the monotonic background.  The oscillation periods found in our crystals agree with those reported by Sakamoto \cite{Sakamoto1976}. The product of the oscillation period and thickness ($\Delta B \cdot d = 14.7 \times 10^{-6}$ T $\cdot$ m) combined with Equation \ref{2} implies $\frac{\partial A}{\partial k_{z}}$ = 1.28 \AA$^{-1}$. Sakamoto showed that this value closely matches an extremum in the belly orbit of the well-established Fermi surface of copper \cite{Sakamoto1976}.  The optimal scaling for the amplitude was found to be $\propto$ $B^{-2.5}$, leading to a nearly constant amplitude for $\delta \sigma B^{2.5}$ over the first few oscillations.  
As discussed above, an exponent of 2.5 is indeed what is expected  for an inflection point singularity of $\frac{\partial A}{\partial k_{z}}$ \cite{gurevich1959oscillations,hurd2012hall, abrikosov2017fundamentals}.

The remarkable difference between the quality of the experimental data in copper and in cadmium is partially an issue of materials science. However, there are fundamental distinctions. There is no exponential term in the field dependence of $\delta \sigma^{pp}$. The thickness dependence is much weaker and, finally, the oscillation amplitude of Hall conductivity is less than half that of the longitudinal conductivity.  All these features are in striking contrast with those observed in cadmium, despite the fact that in both cases, the period of oscillations multiplied by thickness is equal to what is expected by equation \ref{2} and a singular value of $\frac{\partial A}{\partial k_{z}}$.

This leads us to conclude that the specific electronic band structure in Cd is crucial for the emergence of the empirical scaling identified in this study. In contrast to copper, cadmium is a compensated metal with equal volumes of hole-like and electron-like Fermi surface pockets. This may play a role in locking the amplitude of $\delta \sigma_{xy}$ to $\sim 2\delta \sigma_{xx}$. More importantly, in cadmium, $\frac{\partial A}{\partial k_{z}}$ is constant over a large portion of the Fermi surface. In copper, its singularity is sharply localized and corresponds to a single inflection point in momentum space.

\subsection{Interplay between two distinct energy quantizations} 
\begin{figure*}[ht]
\includegraphics[width=0.9\linewidth]{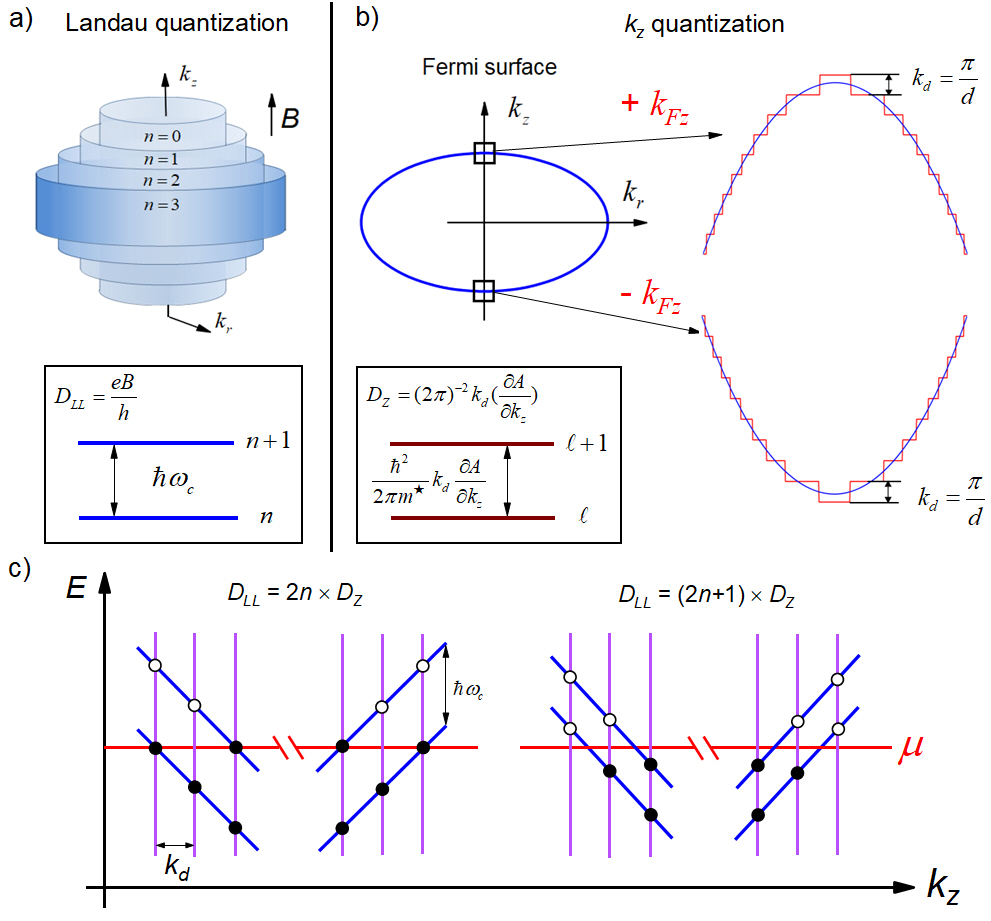} 
\caption{\textbf{Landau tubes and Fermi surface steps.} 
\textbf{(a)} A magnetic field truncates the Fermi surface into Landau tubes.  Neighboring Landau levels, indexed $n$ and $n+1$, are separated by an energy interval, which depends on magnetic field ($\hbar \omega_c$). \textbf{(b)} Confinement of electrons in real space modifies the smooth Fermi surface and induces steps in $k$-space (red line). Their height along $k_z$ is $k_d=\frac{\pi}{d}$. The energy distance between two neighboring levels, indexed $l$ and $l+1$, depends on the thickness $d$, ($\frac{\hbar^2}{2\pi m^{\star}} k_d \partial A/\partial k_{z}$). \textbf{(c)} The vertical lines in the energy, $k_z$ plane refer to the discretization introduced by spatial confinement. Their intersections with Landau levels (oblique lines) are marked as circles (black when occupied and empty when unoccupied). These correspond to eigen-states common to both components of the global Hamiltonian. When the ratio of the two degeneracies per unit square in $k$-space is an even number (left panel), all highest occupied  states have an identical energy and the chemical potential is easily defined. When the ratio of the two DOS is an odd number (right panel), the chemical potential resides at halfway between occupied and unoccupied allowed states. Alternation between these two situations occurs with a periodicity equal to the one found by the semiclassical picture.} 
\label{fig.8}
\end{figure*}

Spatial confinement is known to generate discrete energy levels. Quantum effects are expected when the thickness approaches the Fermi wavelength ($d \approx \lambda_F$). However, our case is far from this limit.  In a finite magnetic field, energy levels become discrete. The semiclassical picture breaks down at high magnetic fields (when $\ell_B \approx \lambda_F$). This is not our case either. 

However, when well-defined $\frac{\partial A}{\partial k_{z}}$ is shared by a significant number of electronic states, a bridge builds up between these two distinct discretizations. 
A magnetic field truncates the Fermi surface into Landau tubes oriented along the  magnetic field (Fig. \ref{fig.8}\textbf{a}). Each tube corresponds to states sharing an identical Landau level index, $n$. Two neighboring Landau levels, $n$ and $n+1$, are separated by an energy of $\hbar \omega_c$, where $\omega_c=\frac{eB}{m^\star}$ is the cyclotron energy and $m^\star$ is the effective mass \cite{Shoenberg_1984}.

A finite thickness of $d$ along the $z$ direction in real space implies that the wave-vector along this direction ceases to be continuous. It can only be an integer times $\frac{\pi}{d}$. Such a constraint sculpts $k_z$ steps on the Fermi surface with a height of $k_d=\frac{\pi}{d}$, as illustrated in Fig. \ref{fig.8}\textbf{b}. Note that $k_d \ell \geq 1$ and therefore, $k_d$ is well-defined. This reorganization redistributes the in-plane and the out-of-plane components of the kinetic energy. In an infinite sample,  $k_r=0$ states have no in-plane kinetic energy and the share of in-plane (out-of-plane) kinetic energy increases (decreases) smoothly with increasing  $k_r$.  In contrast, in a finite sample, $k_d$ steps on the Fermi surface are concomitant with abrupt transfers between the in-plane and out-of-plane components equal to:
\begin{equation}
\delta E= \frac{\hbar^2}{2\pi m^\star} k_d \frac{\partial A}{\partial k_{z}}
\label{deltaE}
\end{equation}

Thus, at zero magnetic field, the out-of-plane and the in-plane kinetic energies are discrete and there are $\delta E$ steps. Now, a finite  magnetic field opens a  cyclotron gap equal to $\hbar \omega_c =\hbar\frac{eB}{m^\star}$ in the in-plane energy spectrum. The interplay between these two distinct energy quantizations, one set by thickness and another by the magnetic field, can provide a straightforward account of oscillations.

Let us consider the degeneracy of states for each type of quantization. The degeneracy of Landau levels per unit area is simply  $D_{LL}=\frac{eB}{h}$. As for $z$-axis confinement, the flat rings carved on the Fermi surface are each a set of degenerate states. In each ring, indexed  $l$, radial wave-vectors (which are good quantum numbers at zero field)  are distributed within a finite window ($k_{Fr,l+1}<k_{Fr}<k_{Fr,l}$). The area of each ring is equal to: 
\begin{equation}
\pi (k_{Fr,l+1}^2-k_{Fr,l}^2) = k_d\frac{\partial A}{\partial k_z}
\label{ring}
\end{equation} 

Each level $l$ has an out-of-plane Fermi energy of  $E_{F\perp,l}= \delta E |l|$ and a degeneracy of $D_{z}=(2\pi)^{-2}k_d\frac{\partial A}{\partial k_z}$ per unit area.

Thus, each discretization generates a set of degenerate states whose energy is identical, but is distinguished either by their radial wave-vector (in the case of confinement) or by their angular momentum (in case of Landau quantization). Let us now introduce a dimensionless quantity, $\Omega$: 
\begin{equation}
\Omega \equiv d\ell_B^{-2}(\frac{\partial A}{\partial k_z})^{-1}
\label{period}
\end{equation} 

One can see that $\Omega$, which is proportional to the magnetic field is also set by the ratio of the two degeneracies:
\begin{equation}
\frac{D_{LL}}{D_{z}} \equiv 2\Omega
\label{degeneracy}
\end{equation} 

The $D_{LL}/D_{z}$ ratio, which quantifies the number of flat rings in each Landau level, is akin to an inverted filling factor. Recall the filling factor in a two-dimensional electron gas, $\nu=\frac{h n_{2d}}{eB}$, which is proportional to the inverse of the magnetic field \cite{yoshioka2002quantum}.  In our case, the number of rings in each Landau level increases linearly with magnetic field.  In both cases, oscillations arise because $\nu$ and $\Omega$ become integers at regular intervals (either of the magnetic field or the inverse of it).  

Thus, as the magnetic field is swept, the inverted filling factor $\Omega$ steadily increases. At specific magnetic fields, it becomes an integer. This situation is depicted on the left-hand side of Figure \ref{fig.8}c. Each discretization is represented by a set of solid lines and states allowed by their combination by empty circles. Commensurability between $D_{LL}$ and $D_z$ allows occupation of allowed states up to the chemical potential. However, this is not the case in an arbitrary magnetic field. The right hand of Figure \ref{fig.8}c  sketches a situation in which $\Omega$ is a half-integer and the chemical potential resides at midway between allowed states. Since dimensionless $\Omega$ periodically becomes an integer with sweeping magnetic field, this picture implies the existence of oscillations that are periodic in $B$, with a period identical to the semiclassical one. Indeed, equation \ref{period} combined with equation \ref{2} yields:
\begin{equation}
\Omega = \frac{B}{\Delta B}
\label{period-2}
\end{equation} 

\begin{figure*}[hbt!]
\includegraphics[width=0.9\linewidth]{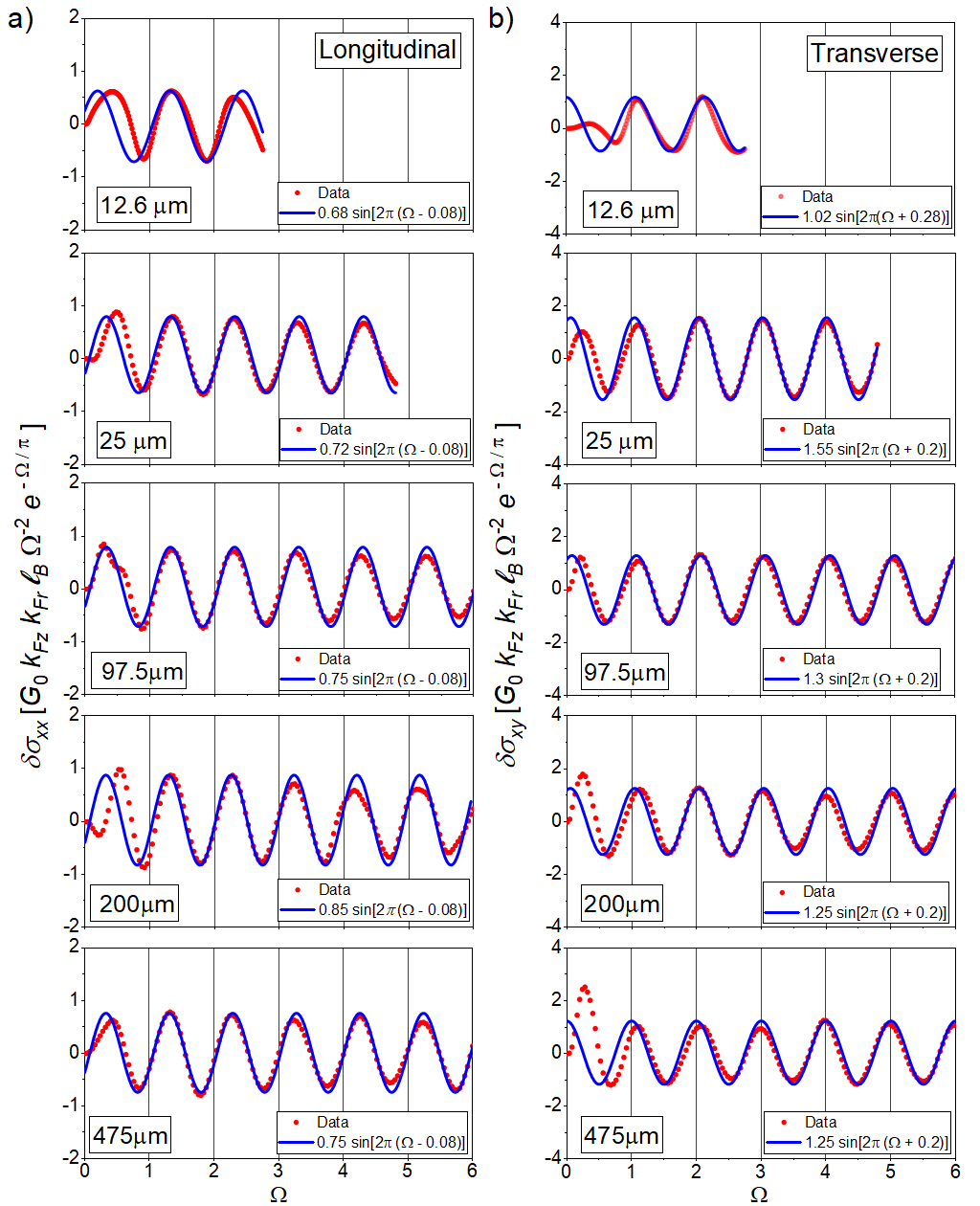} 
\caption{\textbf{Equation \ref{QUSO} and experimental data.} 
\textbf{(a)} Oscillations of dimensionless conductivity as a function of dimensionless period index. The vertical axis is the oscillating conductivity divided by the product of $G_0 = \frac{2e^2}{h}$ and $k_{Fr}k_{Fz}\ell_B\Omega^{-2}e^{-\Omega /\pi}$. The horizontal axis represents  $\Omega=\ell_B^{-2} \left( \frac{\partial A}{\partial k_z}\right)^{-1}d$. The solid blue lines represents $\alpha sin (\Omega - \phi)$. Save for the first oscillation, in all samples, the data matches the sine fit, with $\alpha \approx 1$ and $\phi \approx 0.15 \pi $. \textbf{(b)} Same for Hall conductivity. A similar fit matches the data with an $\alpha$, which is twice larger and a $\phi$, which is shifted by $-\pi/2$.} 
\label{fig.9}
\end{figure*}

Thus, in regular intervals of increasing magnetic field, $\Omega$ becomes an integer and the two quantizations become commensurate. In other words, the boundary between occupied and unoccupied states of the two sets of discrete levels match each other. In contrast, at halfway between two successive integer values of $\Omega$, the chemical potential is located between the energy levels allowed by the two discretizations.  This leads to a competition between the two energy spacings. When the magnetic field is large or the sample is thick, $\hbar \omega_c > \delta E$ and the cyclotron gap dominates. Conversely, when the field is weak or the sample is thin, what dominates is the discontinuity in the out-of-plane kinetic energy imposed by spatial confinement.

Cadmium is a compensated metal. Charge conservation implies that the densities of mobile electrons and mobile holes remain identical in a stoichiometric crystal. However, the chemical potential and the absolute values of carrier density can both shift if it lowers energy. This feature has been documented in detail in the extreme quantum limit of bismuth subject to a strong magnetic field \cite{Zhu2017}. 

In the case depicted in the left side of Fig. \ref{fig.8}c, the occupied and unoccupied states are separated by a cyclotron gap. Tunneling is expected to occur between two neighboring Landau levels. Our empirical equation \ref{OSC} includes a $\exp(-B/B_0)$ term, where $B_0\approx 3\Delta B$. As seen in Figure \ref{fig.5}b, our experimental data can agree with $B_0\simeq \pi \Delta B$ and therefore the exponential term can be $\exp(-\Omega/\pi)$. This leads to the following expression for tunneling action $s$ in $\exp(-s/\hbar)$:
\begin{equation}
s = e B k_d^{-1} (\frac{\partial A}{\partial k_z})^{-1}
\label{tunneling}
\end{equation} 

There is a deep connection between quantum tunneling and the uncertainty principle \cite{razavy2003quantum}. Despite conservation of energy, a particle can tunnel across an energy barrier in real space with a height larger than its kinetic energy. This is because one cannot know both the position and the momentum with infinite accuracy. In the Wentzel-Kramers-Brillouin (WKB) approximation, the tunneling action for a particle with momentum $p$ is given by \cite{razavy2003quantum}: 
\begin{equation}
s = \int_{x_1}^{x_2}p(x')dx'
\label{tunneling-WKB}
\end{equation} 

Comparing Equations \ref{tunneling} and \ref{tunneling-WKB}, one can see that in our case, the action, instead of being a product of momentum and position, is the product of the charge of electron, $e$, and a magnetic flux,  $B k_d^{-1} (\frac{\partial A}{\partial k_z})^{-1}$. Let us recall that Noether's theorem links the conservation of electric charge to the gauge symmetry. In quantum mechanics, the electric charge and magnetic flux are conjugate variables.

An intuitive account of Equation \ref{tunneling} is provided by noting that an electron which does not belong with a Landau level is only allowed because the product of the uncertainties in electric charge and magnetic flux is bounded by the Planck constant. Thus, there is an analogy between such an electron tunneling across two neighboring Landau levels and a particle crossing an energy barrier.
 
 \subsection{The quantum origin of the empirical scaling}

Given the considerations detailed above, we can rewrite Equation \ref{OSC} as:
\begin{equation}
\delta \sigma^{pp} = \alpha G_0 k_{Fz}k_{Fr}\ell_B \Omega^{-2}e^{-\Omega/\pi}
    \label{QUSO}
\end{equation}

Here, $\alpha$ is a dimensionless fitting parameter and $k_{Fz}$, $k_{Fr}$ are the Fermi radii. 

Fig. \ref{fig.9}, confronts our experimental data with equation \ref{QUSO}. For all samples, the normalized $\delta \sigma$ is plotted against the normalized magnetic field.  Both axes are dimensionless. The horizontal axis is $\Omega$, the magnetic field normalized by the [thickness-dependent] period. The vertical axis is $\delta \sigma_{ii}$ or $\delta \sigma_{ij}$ normalized by the conductance quantum and the product of $k_{Fz}k_{Fr}\ell_B \Omega^{-2}e^{-\Omega/\pi}$. One can see that our data fit a sinusoidal function whose amplitude is accurately given by Equation \ref{QUSO} with $\alpha \simeq 1$ and known values of $k_{Fz}$ and $k_{Fr}$. 

A rigorous explanation of the striking success of equation \ref{QUSO} is a task for future studies. Let us make, however, a few brief remarks.

Equation \ref{QUSO} is reminiscent of a three-dimensional Drude conductivity expressed as Landauer transmission ($\sigma=\alpha G_0 k_F^2 \ell $).  A naive interpretation would be the following: As the magnetic field is swept, a fraction ($\propto \Omega^{-2}e^{-\frac{\Omega}{\pi}}$) of states at Fermi level oscillate between localization and conduction. When they conduct their mean free path, is  $\approx \ell_B$.

The exponential term in Equation \ref{QUSO}, which refers to tunneling across neighboring Landau levels, is to be contrasted with another case of tunneling, namely magnetic breakdown \cite{Shoenberg_1984}. There, tunneling occurs across cyclotron orbits separated by an energy barrier. We also note that tunneling between Landau levels across an $n-p$ junction of Weyl semi-metals was framed \cite{Saykin2018} as a manifestation of Landau-Zener tunneling in a two-level system \cite{Kitamura2020}. A similar framework was proposed for the case of magnetic breakdown \cite{Alexa2017}. The two cases give rise to an exponential term with magnetic field either in the numerator or the denominator of the exponent. The relevance of these ideas to our experimental context remains an open question.

A comparison of oscillations in the transverse and in the longitudinal channels reveals puzzles that are yet to be understood and constitute a  challenge to theory. In all five samples studied, there is a $\pi/2$ phase shift between the two sets of oscillations. Moreover, the transverse amplitude is roughly twice the longitudinal amplitude (see Fig. \ref{fig.9} and  the supplementary note 6\cite{SM} for more details).

\section{Conclusion}
In summary, we showed that in thin single crystals of cadmium hosting ballistic electrons, Sondheimer (or magnetomorphic)  oscillations of conductivity display a field dependence never encountered before.  Their amplitude is set by the quantum of conductance and a length scale depending only on the thickness, the magnetic length, and the Fermi surface geometry.  We argued that to explain this experimental observation, one needs to go beyond the available semiclassical theories and take into account the interplay between Landau tubes and the steps sculpted on the Fermi surface by spatial confinement. We then showed this allows us to understand the empirical scaling as a simple expression in which the key parameter is a dimensionless parameter set by the Fermi surface geometry.  

\section{Data availability}
The data that support the findings of this study are available from the corresponding author upon reasonable request.

\section{Acknowledgments}
We are grateful to  Benoît Fauqu\'e, Mikhail Feigel'man, Justin Song and Alaska Subedi for discussions. This work is part of a Cai Yuanpei Franco-Chinese collaboration program (No. 51258NK).  In China, it was supported by the National Science Foundation of China (Grant No. 12474043, 12004123 and 51861135104), the National Key Research and Development Program of China (Grant No. 2022YFA1403500), and the Fundamental Research Funds for the Central Universities (Grant no. 2019kfyXMBZ071). X. L. was supported by The National Key Research and Development Program of China (2023YFA1609600) , the National Natural Science Foundation of China (12304065) and the Hubei Provincial Natural Science Foundation ‌(2025AFA072).  We also acknowledge a grant from the \^Ile de France region.
\section{Author contributions}
X.G., with assistance from X.L., performed the FIB and low-temperature transport measurements. L.Z. grew the cadmium samples. Z.Z. and K.B. conceived the project. X.G., Z.Z., and K.B. analyzed the data and wrote the manuscript, with contributions from all other authors.

\section{Competing interests}
The authors declare no competing interests.

\clearpage

\begin{center}{\large\bf Supplementary Materials for ``Scalable Sondheimer oscillations driven by commensurability between two
quantizations"}\\
\end{center}

\renewcommand{\thesection}{S\arabic{section}}
\renewcommand{\thetable}{S\arabic{table}}
\renewcommand{\thefigure}{S\arabic{figure}}
\renewcommand{\theequation}{S\arabic{equation}}

\setcounter{section}{0}
\setcounter{figure}{0}
\setcounter{table}{0}
\setcounter{equation}{0}

\section*{Note 1. Materials and Methods.}
\begin{figure*}[bht!]
\includegraphics[width=0.8\linewidth]{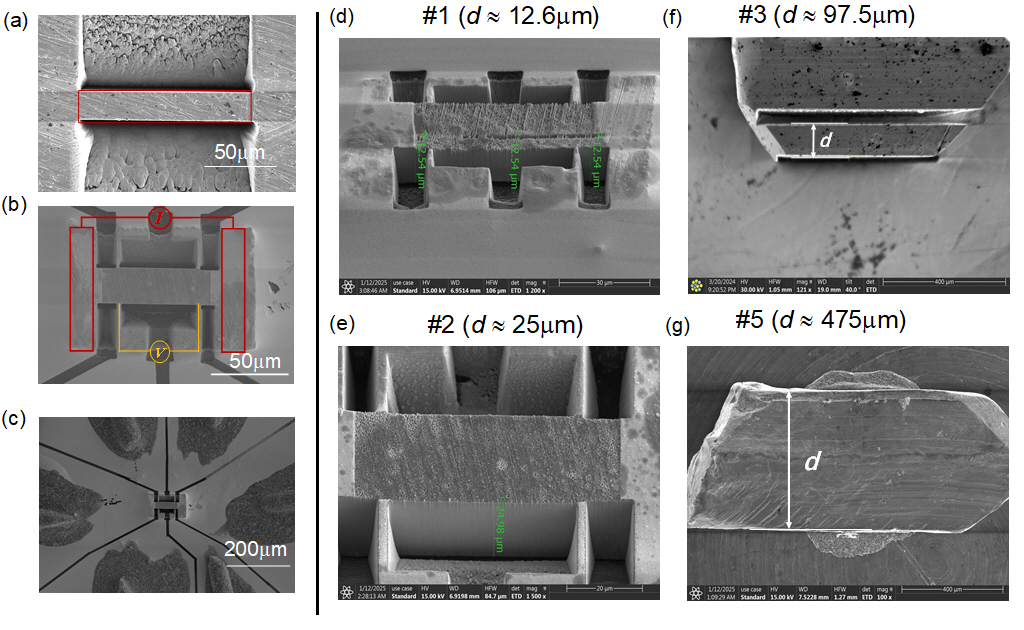} 
\centering
\caption{\textbf{Sample preparation and thickness measurement.} 
\textbf{(a)} Scanning electron-beam microscopy (SEM) image of the sample (marked by red box) etched by FIB. The scale bar denotes 50 $\mu$m. \textbf{(b)} Ion beam etching was employed to fabricate electrodes for the samples. The current electrode is highlighted with the red box, and voltage electrodes are marked by yellow line.  \textbf{(c)} Attach the conductive electrode to the gold wire with silver gel. Thickness measurements were conducted on samples of 12.6 $\mu$m \textbf{(d)}, 25 $\mu$m \textbf{(e)}, 97.5 $\mu$m \textbf{(f)} and 475 $\mu$m \textbf{(g)}.
}
\label{fig.S1}
\end{figure*}

\textbf{Crystal growth.} Cd single crystals were grown using the vapor-phase transport method. A quartz tube containing appropriate amounts of starting material, Cd (99.9997\%), was kept at the growth temperature for two weeks in a two-zone furnace. Shiny plate-like crystals were produced under a temperature gradient of 300 –200$^{\circ}$C.  

\textbf{Fabrication of the samples.} The production method followed previous work \cite{van2021sondheimer}. First, the crystallographic orientation of the bulk crystal was determined using X-ray diffraction. Subsequently, the crystal was securely affixed to an SEM stub in the desired orientation. The next critical step was to employ a focused beam of Ga ions to meticulously carve a rectangular slab, a lamella, from the crystal. This process unfolds in three stages.

\begin{figure*}[bht!]
\includegraphics[width=0.9\linewidth]{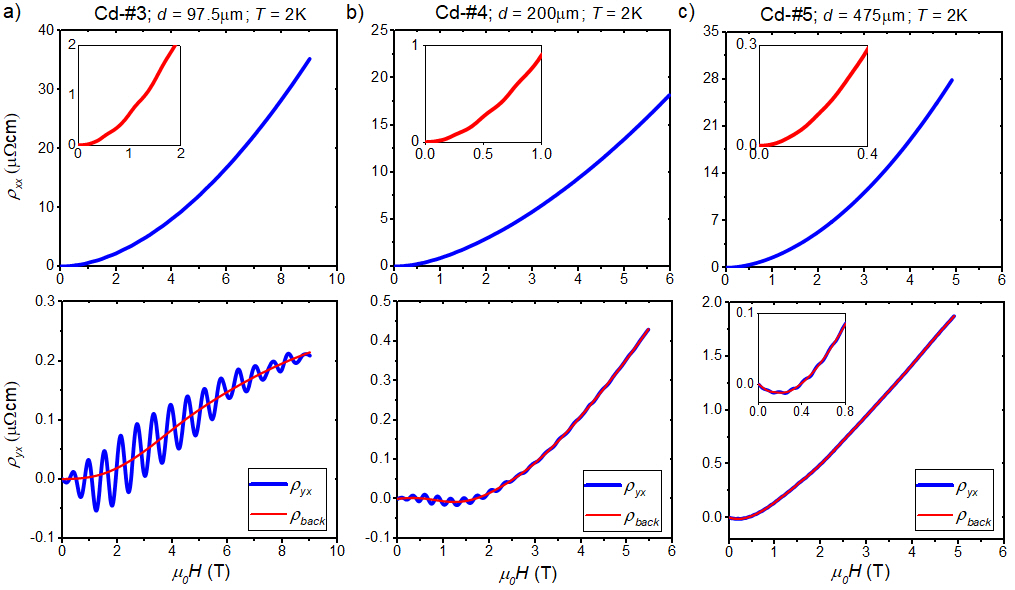} 
\caption{\textbf{Longitudinal and transverse resistivity  in the thicker samples.} 
 Longitudinal and transverse resistivity at 2 K in the 97.5$\mu$m \textbf{(a)},  in the 200$\mu$m  \textbf{(b)} and  in the 475$\mu$m  \textbf{(c)} samples.
}
\label{fig.S2}
\end{figure*}

\begin{figure*}[bht!]
\includegraphics[width=1.0\linewidth]{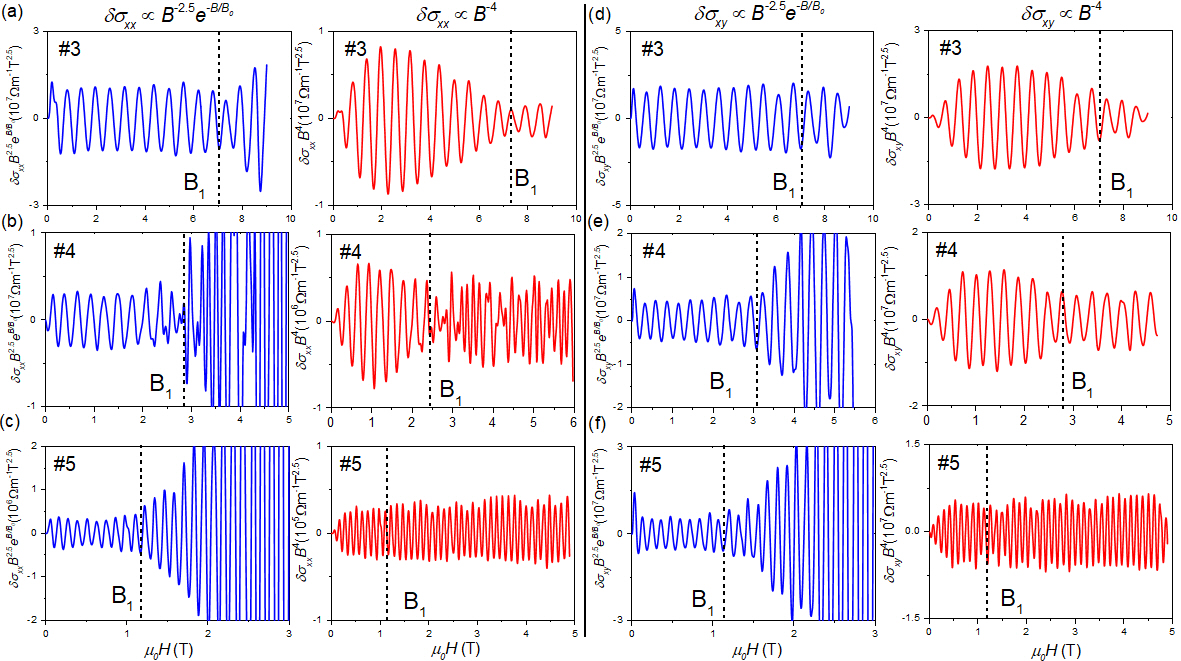} 
\caption{\textbf{Field dependence of the amplitude of oscillations in the thicker samples.} 
\textbf{(a)} The oscillating longitudinal conductivity, $\delta \sigma_{xx}$ multiplied by $B^{2.5} e^{B/B_0}$ (left) and by $B^{4}$ (right) in the 97.5 $\mu$m sample. \textbf{(b)} Same for the 200 $\mu$m sample. \textbf{(c)} Same for the 475 $\mu$m sample. \textbf{(d)}, \textbf{(e)} and \textbf{(f)} Same for oscillating Hall conductivity $\delta \sigma_{xy}$. In all cases, below $B_1$, the first multiplication yields a flat amplitude but not the second. Above $B_1$, it is the inverse.}
\label{fig.S3}
\end{figure*}

Initially, a high current of 50 nA was used to create two gaps within the crystal (Fig.\ref{fig.S1}\textbf{a}). These gaps were separated by a section of the crystal, approximately $\sim$25 $\mu$m thick and 150 $\mu$m long. Following this, a smaller current of 15 nA was applied to smoothen the crystal section, resulting in moderately flat sidewalls. The lamella remained attached to the parent crystal through two beams at its top.

Afterwards, using a precise 7 nA current, both sides of the lamella were cut, ensuring parallel sidewalls and a high level of smoothness. The obtained flakes were then carefully transferred to an alumina substrate. It is crucial to note that a very thin layer of ``Araldite'' epoxy adhesive must be applied to the substrate beforehand. Once positioned atop the epoxy adhesive, capillary forces naturally shaped the epoxy around the lamella, extending smoothly to each of its top surfaces without covering them.

After allowing the epoxy adhesive to dry naturally, the substrate was taken to a sputtering machine, depositing a 300 nm layer of Au on the lamella. To further refine the device, the Scios 2 system was utilized. Initially, the Au layer covering the active part of the device was removed with an acceleration voltage of 5 kV and an ion current of 2 nA. Subsequently, the overall device shape, including the contact positions, was precisely cut out at 0.3 nA and 30 kV. Finally, the Au layer away from the device was severed, ensuring a current flow exclusively through the device. With these steps completed, the device became ready for measurement (Fig.\ref{fig.S1} \textbf{b} and \textbf{c}). 

Samples with a thickness of approximately 12.6 and 25 $\mu$m are  shown in Fig. \ref{fig.S1} \textbf{d} and \textbf{e}. 
We used a focused ion beam to etch lamellae with sufficient size to measure resistance using the four-electrode method, as shown in Fig.\ref{fig.S1} \textbf{f} and \textbf{g}.

\section*{Note 2. Estimation of the mean-free-path in Cd single crystals.}
Cadmium is a compensated metal with a carrier density of $n=p=3.48\times 10^{22}~$cm$^{-3}$ \cite{Subedi2024}. The free-electron Drude conductivity is $\sigma =\frac{2e^2}{3\pi\hbar}(k_{Fe}^2+k_{Fh}^2)\ell_0$. Assuming {$k_{Fe}=k_{Fh}=(3\pi^2n)^{\frac{1}{3}}$, one finds that $\rho_0 \ell=11.5\times 10^{-16}~\Omega\cdot$m$^2$, which is close to what was deduced  previously by Gall \cite{Gall2016}.  This is the expression that we used to extract $\ell_0$   from $\rho_0$ for each sample in table I. As seen in the table, the extracted value is $\approx$ 100 $\mu$m in the thickest sample, and remains comparable to the thickness with thinning.

\section*{Note 3. Magneto-morphic oscillations in  samples with a thickness between  97.5$\mu$m to 475$\mu$m. }

Fig.\ref{fig.S2} shows magneto-morphic oscillations of longitudinal and transverse electrical resistivity in the three thickest cadmium single crystals at 2 K. Sondheimer oscillations are clearly visible in the Hall signal, where the background is smaller due to the compensation between electrons and holes. By inverting the measured resistivity tensor, we quantified the conductivity tensor using $\sigma_{ij} = \frac{-\rho_{ij}}{\rho_{ii}^2 + \rho_{ij}^2}$. 

As shown in Fig.\ref{fig.S3} \textbf{a}, \textbf{b}, and \textbf{c}, when $B<B_1$, the oscillating longitudinal conductivity scales as  $\propto$ $B^{-2.5} e^{-B/3\Delta B}$. When $B>B_1$, its amplitude follows $B^{-4}$. The same result is obtained for the oscillating Hall conductivity, as shown in Fig.\ref{fig.S3} \textbf{d}, \textbf{e}, and \textbf{f}.

\section*{Note 4. The field and thickness dependence of oscillations in the two regimes}

\begin{figure*}[bht!]
\includegraphics[width=0.75\linewidth]{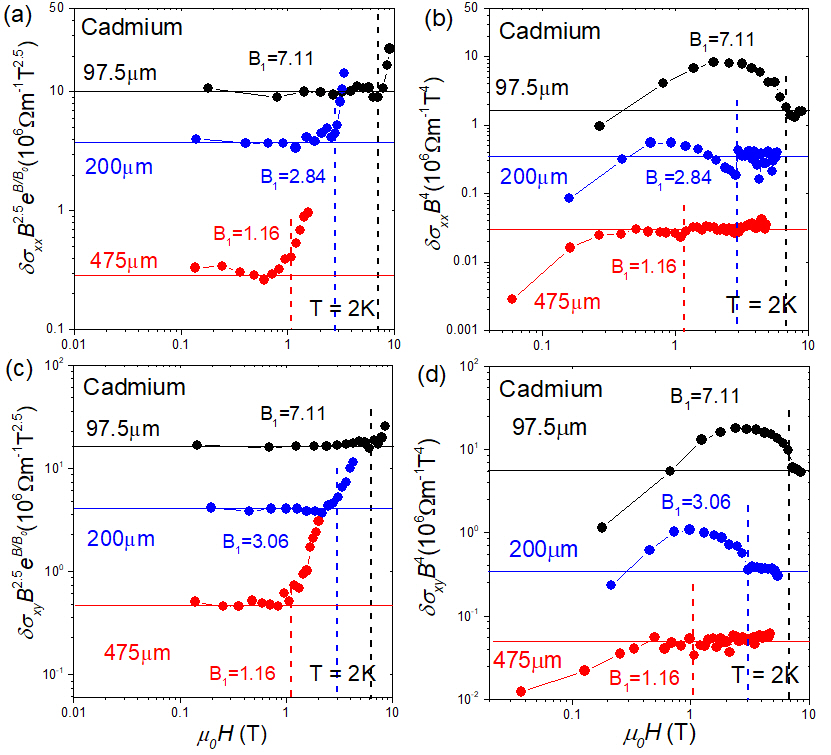} 
\caption{\textbf{The field dependence of oscillation amplitude in two regimes.} 
(a) Longitudinal conductivity multiplied  by  $\propto B^{2.5} e^{B/B_0}$ as function of magnetic field. (b) Longitudinal conductivity multiplied  by  $\propto B^{4}$ as function of magnetic field. The flatness in two different field windows implies that when the field is less than $B_1$, the SO oscillation can be expressed as $\delta\sigma_{xx} \propto B^{-2.5} e^{-B/B_0}$, but when the field is larger than $B_1$, it follows $B^{-4}$. (c), (d) Same for Hall conductivity.}
\label{fig.S4}
\end{figure*}

As mentioned above, $B_1$ is the threshold field separating the two types of field-dependent amplitude. When $B < B_1$, multiplying the amplitude of oscillations by $d^2$ and plotting $\delta \sigma_{ij} B^{2.5} d^2$ yields the results shown in Fig. \ref{fig.S5}\textbf{a} and \textbf{b}, in which all curves are on top of each other.

\begin{figure*}[bht!]
\includegraphics[width=0.85\linewidth]{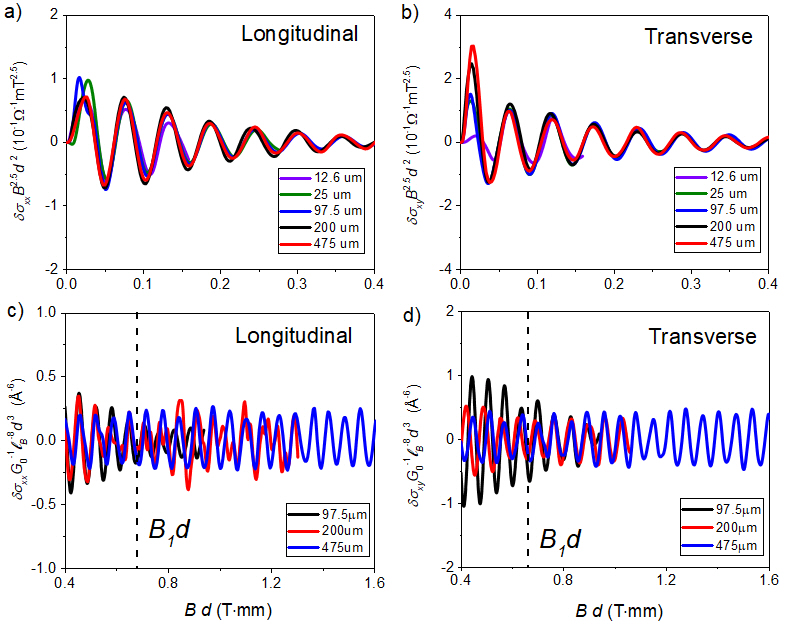} 
\caption{\textbf{Scaling of the oscillations with thickness.} When $B < B_1$, \textbf{(a)} the amplitude of the $\sigma_{xx}$ oscillation multiplied by $d^2$ is plotted as a function of the product of the field and thickness. All curves collapse. \textbf{(b)} Same procedure for the Hall conductivity $\sigma_{xy}$. When $B > B_1$, \textbf{(c)} the vertical axis is the conductivity divided by the quantum of conductance $G_0$ and multiplied by $\ell_B^{-8} \times d^3$. The horizontal axis, linear in magnetic field, represents the dimensionless product: $B d$. \textbf{(d)} Same for the Hall conductivity $\sigma_{xy}$. }
\label{fig.S5}
\end{figure*}

On the other hand, when the field exceeds $B_1$, the amplitude falls off as $\propto B^{-4}$. Multiplying the amplitude of oscillations by $d^3$  and plotting $\delta \sigma_{ij} G_0^{-1}\ell_{B}^{-8} d^3$ leads to Fig. \ref{fig.S5}\textbf{c} and \textbf{d}. One can see that the scaling is not as good as in the case of low-field ($B<B_1$) oscillations.  Nevertheless,  $d^{-3}$ gives a much better scaling than $d^ {-2}$ and $d^ {-4}$, indicating that the thickness dependence in this regime is close to  $d^{-3}$.


\section*{Note 5. Magneto-morphic oscillations in copper samples.}

Fig.\ref{fig.S6} shows magneto-morphic oscillations in longitudinal and transverse electrical resistivity for copper single crystals at 2 K. Inverting the measured resistivity tensor, we quantified the conductivity tensor using $\sigma_{ij} = \frac{-\rho_{ij}}{\rho_{ii}^2 + \rho_{ij}^2}$. The non-oscillatory background is obtained through polynomial fitting (red line in Fig. \ref{fig.S6}\textbf{c}) . After subtracting this background, the oscillating component was extracted. We found that multiplying the extracted signal by $B^{2.5}$ yields an almost constant amplitude for the first few oscillations, as shown in Fig. \ref{fig.S6}\textbf{d} (top panel).

Consistent with this observation, a similar scaling ($\propto B^{2.5}$) is observed in the Hall conductivity of the same crystal, as shown in Fig. \ref{fig.S6}\textbf{d} (bottom panel). Applying the same method to the 200 $\mu$m sample, the field dependence for both the longitudinal conductivity and transverse conductivity was found to scale as $\propto B^{2.5}$, as shown in Fig. \ref{fig.S6}\textbf{e}.

Fig. \ref{fig.S6}\textbf{f} shown the oscillation period as a function of thickness $d$. The product of the oscillation period and thickness, $\Delta B \times d$, is equal to 14.7 $\times$ $10^{-6}$ T $\cdot$ m, which is same with previous report in copper crystal\cite{Sakamoto1976}.


\begin{figure*}[ht]
\includegraphics[width=0.9\linewidth]{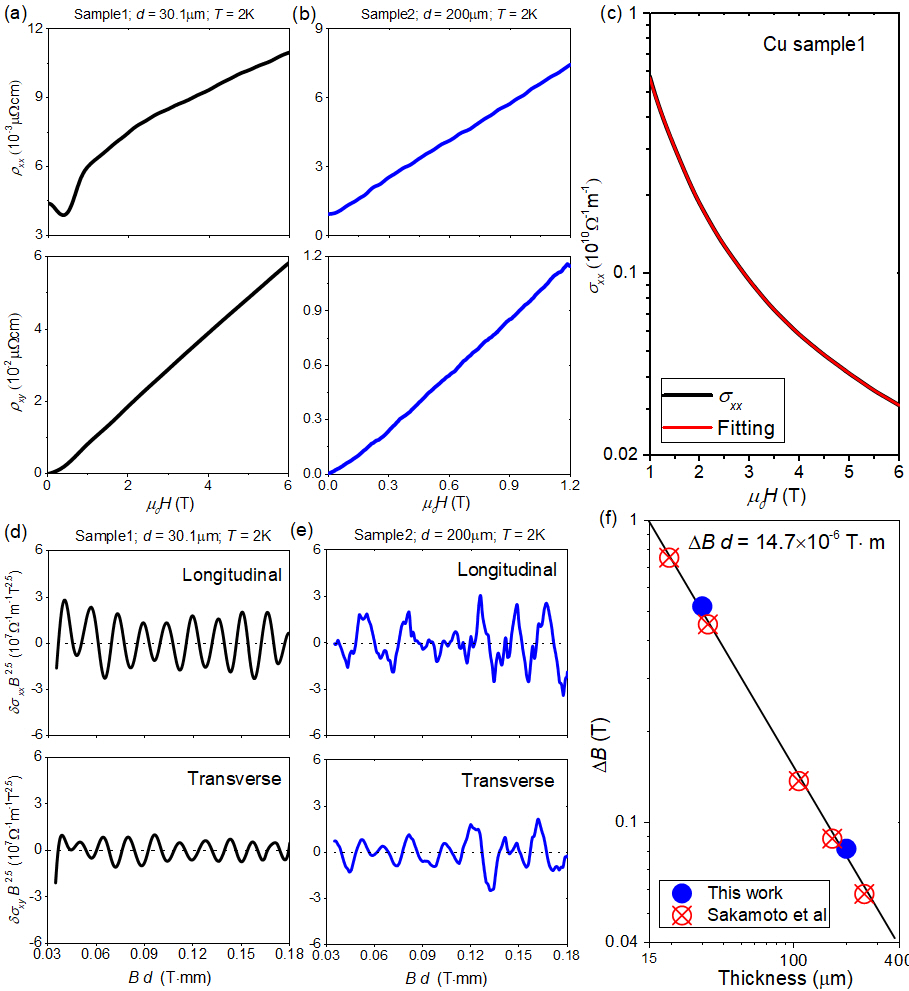} 
\caption{\textbf{Sondheimer oscillations in copper crystals.}  
\textbf{(a)} Longitudinal (top) and transverse (bottom) resistivity in the 30.1 $\mu$m sample. Sondheimer oscillations can be observed in the longitudinal resistivity.   \textbf{(b)} Same for the 200 $\mu$m sample. \textbf{(c)} Longitudinal conductivity in 30.1 $\mu$m sample. The red line represents a polynomial fit to the background. \textbf{(d)} Oscillating component of conductivity multiplied by $B^{2.5}$ as function of the product of magnetic field and thickness ($B d$) in a copper crystal with a thickness of 30.1 $\mu$m. \textbf{(e)} Oscillating component of conductivity multiplied by $B^{2.5}$ as function of $B d$  in another copper crystal with a thickness of 200 $\mu$m. All data were taken at $T$ = 2 K with the field applied along the [100] orientation. \textbf{(f)} The variation of the period of oscillation with thickness in our two samples together with those reported in a previous report \cite{Sakamoto1976}. } 
\label{fig.S6}
\end{figure*}




\section*{Note 6. Oscillations of dimensionless conductivity as a function of dimensionless period.}

As mentioned in the main text, we can rewrite the empirical expression \ref{OSC} using $\Omega$ as: $\delta \sigma^{pp} = \alpha G_0 k_{Fz} k_{Fr} \ell_B \Omega^{-2} e^{-\Omega/\pi}$, where $\alpha$ is a dimensionless fitting parameter.

A sine fit, $\alpha$sin$(\Omega - \phi)$, is applied to the data of the 12.6, 25, 97.5, 200 and 475 $\mu m$ samples, marked by a solid blue line in the main text Fig. \ref{fig.9}. The fitting amplitude $\alpha$ are shown in Fig. \ref{fig.S7}\textbf{a}, revealing that for longitudinal conductivity, $\alpha \approx 1$, while for transverse conductivity, $\alpha$ is approximately twice the longitudinal value, as shown in Fig. \ref{fig.S7}\textbf{b}. Fig. \ref{fig.S7}\textbf{c} show the phase in longitudinal and transverse conductivity, respectively. The phase difference, $\phi_{xy}$ $-$ $\phi_{xx}$, is close to $\frac{- \pi}{2}$, as shown in Fig. \ref{fig.S7}\textbf{d}. 


\begin{figure*}[ht]
\includegraphics[width=0.8\linewidth]{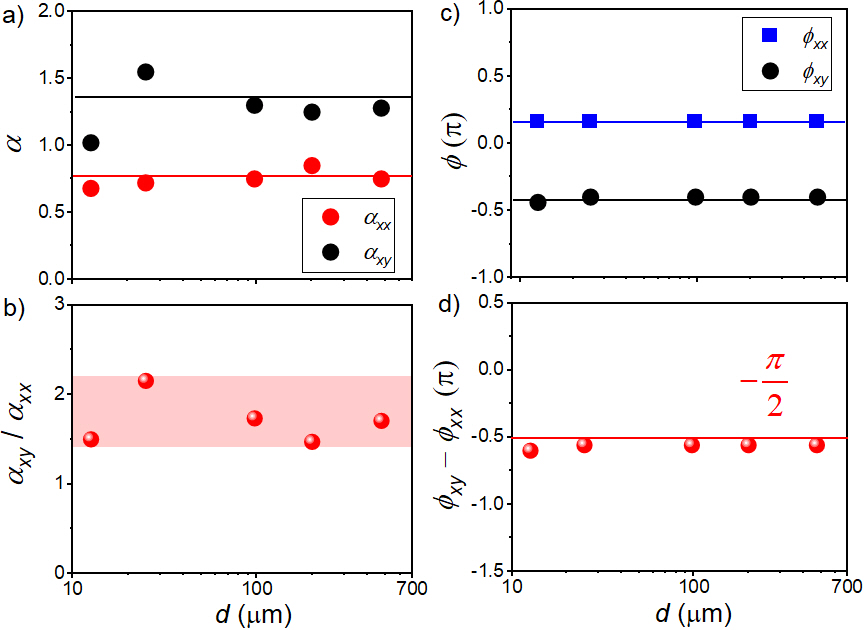} 
\caption{\textbf{Amplitude and phase of oscillations.} Fitting our data to $\alpha sin (\Omega - \phi)$ for samples with different thickness (Fig.\ref{fig.9} in the main text) yields an $\alpha$ (panel a) and a $\phi$  (panel c) for each sample. \textbf{(b)} Ratio of the longitudinal and transverse $\alpha$  is close to two in all samples. \textbf{(d)} The shift between $\phi$ in longitudinal and transverse conductivity is close to $-\frac{\pi}{2}$.}
\label{fig.S7}
\end{figure*}

\section*{Note 7. Derivation of energy distance between two eighboring levels $l$ and $l+1$.}

The main text shows (Fig.\ref{fig.8}\textbf{b}) that the confinement of electrons in real space modifies the smooth Fermi surface and induces a stepped structure in $k$-space. The energy separation $\delta E$ between two neighboring levels, $l$ and $l+1$, can be expressed as:
\begin{equation}
\begin{aligned}
\delta E &= \frac{\partial E}{\partial k_z}\Delta k_z \\
&= \frac{\hbar^2}{2m^{\star}}\frac{\partial k^2 }{\partial k_z}k_d \\
& = \frac{\hbar^2}{2\pi m^{\star}} k_d \frac{\partial A } {\partial k_z}
\end{aligned}
\label{EQS1}
\end{equation}
This result corresponds to Eq. 7 in the main text. Furthermore, for a given region of phase space, the degeneracy is defined as the number of quantum states it contains, which is proportional to the ratio of the region's volume to the volume of a single quantum state. Therefore, the degeneracy within the ring of area $k_d \frac{\partial A}{\partial k_z} $ is:
\begin{equation}
D_{z} = \frac{k_d \frac{\partial A}{\partial k_z}}{(2\pi)^2} =(2\pi)^{-2}k_d\frac{\partial A}{\partial k_z}
\label{EQS2}
\end{equation}

\end{document}